\documentclass[11pt,onecolumn]{IEEEtran}
\usepackage{amsthm}
\usepackage{times,amssymb,amsmath,amsfonts,float,nicefrac,color,bbm,mathrsfs,float}
\usepackage{enumerate,multirow,caption,tikz,graphicx}
\usepackage{algpseudocode}
\usepackage[noadjust]{cite}
\usepackage{subcaption}
\usepackage{diagbox}
\usepackage[ruled,vlined,linesnumbered]{algorithm2e}

\usepackage[top=1in,bottom=1in,left=1in,right=1in]{geometry}
\allowdisplaybreaks

\newcommand{\RN}[1]{%
	\textup{\expandafter{\romannumeral#1}}%
}

\usetikzlibrary{shapes.misc, positioning}
\usetikzlibrary{decorations.pathreplacing}

\tikzset{
	block/.style    = {draw, thick, rectangle, minimum width = 1em},
	sblock/.style      = {draw, thick, rectangle, minimum height = 3em,
		minimum width = 3em}, 
}

\tikzset{XOR/.style={draw,circle,append after command={
			[shorten >=\pgflinewidth, shorten <=\pgflinewidth,]
			(\tikzlastnode.north) edge (\tikzlastnode.south)
			(\tikzlastnode.east) edge (\tikzlastnode.west)
		}
	}
}

\newcommand\remove[1]{}

\allowdisplaybreaks

\newtheorem{remark}{Remark}

\newtheorem{cnstr}{Construction}

\def\mathbi#1{{\textbf{\textit #1}}}

\newcommand{\cA}{\mathcal{A}}

\newcommand{\cY}{\mathcal{Y}}

\DeclareMathOperator{\polar}{polar}
\DeclareMathOperator{\RM}{RM}
\DeclareMathOperator{\DP}{DP}
\DeclareMathOperator{\dB}{dB}
\DeclareMathOperator{\wt}{wt}

\DeclareMathOperator{\ML}{ML}

\DeclareMathOperator{\SCL}{SCL}

\DeclareMathOperator{\minus}{minus}

\begin{document}
	\title{A Dynamic Programming Method to Construct Polar Codes with Improved Performance}

	\author{Guodong Li \and  \hspace*{.5in} Min Ye  \and  \hspace*{.5in}  Sihuang Hu}

	\maketitle
	{\renewcommand{\thefootnote}{}\footnotetext{
			
			\vspace{-.2in}
			
			\noindent\rule{1.5in}{.4pt}
			
			Guodong Li is with School of Cyber Science and Technology, Shandong University, Qingdao, Shandong, 266237, China. Email: guodongli@mail.sdu.edu.cn
			
			Min Ye is with Tsinghua-Berkeley Shenzhen Institute, Tsinghua Shenzhen International Graduate School, Shenzhen 518055, China. Email: yeemmi@gmail.com
			
			Sihuang Hu is with  Key Laboratory of Cryptologic Technology and Information Security, Ministry of Education, Shandong University, Qingdao, Shandong, 266237, China and School of Cyber Science and Technology, Shandong University, Qingdao, Shandong, 266237, China. Email: husihuang@sdu.edu.cn
		}
	}

	\renewcommand{\thefootnote}{\arabic{footnote}}
	\setcounter{footnote}{0}

	\begin{abstract}
		In the standard polar code construction, the message vector $(U_0,U_1,\dots,U_{n-1})$ is divided into information bits and frozen bits according to the reliability of each $U_i$ given $(U_0,U_1,\dots,U_{i-1})$ and all the channel outputs.
		While this reliability function is the most suitable measure to choose information bits under the Successive Cancellation (SC) decoder, there is a mismatch between this reliability function and the Successive Cancellation List (SCL) decoder because the SCL decoder also makes use of the information from the future frozen bits.
		
		We propose a Dynamic Programming (DP) construction of polar codes to resolve this mismatch. Our DP construction chooses different sets of information bits for different list sizes in order to optimize the performance of the constructed code under the SCL decoder. Simulation results show that our DP-polar codes consistently demonstrate $0.3$--$1$dB improvement over the standard polar codes under the SCL decoder with list size $32$ for various choices of code lengths and code rates.
	\end{abstract}

	\section{Introduction}\label{sect:intro}
	In Ar{\i}kan's seminal work \cite{Arikan09}, Polar codes were proposed and shown to achieve capacity of any binary-input memoryless symmetric (BMS) channel under the Successive Cancellation (SC) decoder. Since then, various code construction algorithms and decoders have been proposed for polar codes.
	Notable polar code construction algorithms include the channel upgrading/degrading method \cite{Tal13}, the density evolution method \cite{Mori09,Mori09a}, and the  Monte Carlo method proposed by Ar{\i}kan in his original paper \cite{Arikan09}.
	On the decoding side, the Successive Cancellation List (SCL) decoder \cite{Tal15} achieves almost the same performance as the Maximum Likelihood (ML) decoder when the list size is $32$. The introduction of CRC (Cyclic Redundancy Check) in the code construction and the SCL decoder further reduces the decoding error probability of polar codes \cite{Tal15,Niu12}.
	
	In the polar coding scheme, the message vector $(U_0,U_1,\dots,U_{n-1})$ of length $n=2^m$ is encoded into the codeword 
	$
	(X_0,X_1,\dots,X_{n-1})=(U_0,U_1,\dots,U_{n-1})
	\begin{bmatrix}
		1 & 0 \\
		1 & 1
	\end{bmatrix}^{\otimes m} ,
	$
	where $\otimes$ is the Kronecker product. Each $X_i$ is then transmitted through a BMS channel $W$, and the corresponding channel output is denoted as $Y_i$. Commonly-used polar code constructions \cite{Tal13,Mori09,Mori09a,Arikan09} consist of two steps. The first step is to calculate  the reliability of each $U_i$ given $(U_0,U_1,\dots,U_{i-1})$ and $(Y_0,Y_1,\dots,Y_{n-1})$, e.g., the conditional entropy\footnote{Other reliability functions such as the Bhattacharyya parameter are also adopted in some construction algorithms of polar codes, but there are no fundamental differences between these choices of reliability functions.} $H(U_i|U_0,U_1,\dots,U_{i-1},Y_0,Y_1,\dots,Y_{n-1})$. Then the next step is to divide $(U_0,U_1,\dots,U_{n-1})$ into information bits and frozen bits. More precisely, when constructing an $(n,k)$ polar code, we choose top $k$ bits with the highest reliability in $(U_0,U_1,\dots,U_{n-1})$ as the information bits. The other $n-k$ bits are the frozen bits, and we set their values to be $0$.
	While $H(U_i|U_0,U_1,\dots,U_{i-1},Y_0,Y_1,\dots,Y_{n-1})$ serves as the most suitable measure to choose information bits under the SC decoder, there is a mismatch between this reliability function and the SCL decoder. This is because the SC decoder only makes use of the previous frozen bits and the previously decoded information bits while the SCL decoder also makes use of the future frozen bits. Moreover, the SCL decoder is able to use information from more future frozen bits when it has a larger list size. In fact, when the list size is larger than $2^k$, the SCL decoder is the same as the ML decoder. In this case, it can make use of all the future frozen bits to decode the current information bit. Therefore, in order to accurately measure the reliability of $U_i$ in the SCL decoder, one needs to estimate the conditional entropy of $U_i$ given both the previous bits and some of the future frozen bits, where the number of future frozen bits here increases with the list size in the SCL decoder. However, such an estimate is difficult to obtain because the number of future frozen bits whose information can be used when decoding the current bit $U_i$ in the SCL decoder changes with both the list size and the index of $U_i$. As a consequence, we can not follow the two-step procedure in the standard polar code constructions \cite{Tal13,Mori09,Mori09a,Arikan09}. Instead, we propose a new construction method based on dynamic programming to obtain polar codes with better performance under the SCL decoder.
	
	Our new construction makes use of the Plotkin $(\mathbi{u},\mathbi{u}+\mathbi{v})$ decomposition of polar codes. More precisely, for an $(n,k)$ polar code, the vector $\mathbi{u}$ in the Plotkin decomposition is a codeword of a length-$n/2$ polar code, and the vector $\mathbi{v}$ is a codeword of another length-$n/2$ polar code. The dimensions of these two length-$n/2$ polar codes sum up to $k$. Given the list size $L$, our code construction algorithm produces the ``minus" array $\{\minus_L(n,k):n=2^m,1\le m\le 10,~ 0\le k\le n\}$, where $\minus_L(n,k)$ is the number of information bits assigned to the $\mathbi{v}$ branch when the code length is $n$ and code dimension is $k$. In other words, when constructing an $(n,k)$ polar code for the SCL decoder with list size $L$, the $\mathbi{v}$ branch is an $(n/2, \minus_L(n,k))$ polar code, and the $\mathbi{u}$ branch is an $(n/2, k- \minus_L(n,k))$ polar code. By recursively applying the Plotkin decomposition, the ``minus" array allows us to construct polar codes with code length\footnote{Note that our construction algorithm works for any code length if there is enough computing resource. In this paper we limit ourselves to $n\le 1024$ because the construction for larger code length requires more computing resource than we can afford.} $n\le 1024$ and arbitrary code dimension.
	
	The core of our code construction is to calculate the ``minus" array, and this is done by a dynamic programming algorithm. We calculate $\minus_L(n,k)$ from small values of $n$ to large values of $n$. Therefore, when calculating $\minus_L(n,k)$, we already know all the values of $\{\minus_L(n',k'):n'<n,0\le k'\le n'\}$, which allows us to construct polar codes with length $n/2$ and arbitrary code dimension. In order to find the value of $\minus_L(n,k)$, we test all possible choices $k^-$ between $\max(0,k-n/2)$ and $\min(k,n/2)$. When testing the performance of $\minus_L(n,k)=k^-$, we use the minus array $\{\minus_L(n',k'):n'<n,0\le k'\le n'\}$ to construct the two length-$n/2$ polar codes in the $\mathbi{u}$ branch and the $\mathbi{v}$ branch, and then we use the $(\mathbi{u},\mathbi{u}+\mathbi{v})$ construction to combine these two length-$n/2$ polar codes into an $(n,k)$ polar code. The performance of the choice $\minus_L(n,k)=k^-$ is measured by the decoding error probability under the SCL decoder with list size $L$ when transmitting the above $(n,k)$ polar code over a properly chosen testing channel. The decoding error probability is estimated through Monte Carlo simulations, and the value of $\minus_L(n,k)$ is chosen to minimize the decoding error probability. In this way, our new polar code construction is specifically optimized for the SCL decoder with a given list size.
	
	We conduct extensive simulations to compare the performance of our DP-polar codes and the standard polar codes over the binary-input AWGN channel. When we do not use CRC on either code construction, our DP-polar codes consistently demonstrate $0.3$--$1$dB improvement over the standard polar codes under the SCL decoder with list size $32$ for various choices of code rates and code length. When we use CRC on both code constructions, the performance of DP-polar code is similar to the standard polar code construction.
	
	In addition to the improvement on the decoding error probability, our new code construction reveals an unexpected connection between polar codes and Reed-Muller (RM) codes. As mentioned above, our code construction varies with the list size of the SCL decoder. We find that when we set the list size to be $1$, the DP-polar codes are almost the same as the standard polar codes. On the other hand, if we set the list size to be very large, then the DP-polar codes are almost the same as the RM codes. Since the SCL decoder is almost the same as the ML decoder when the list size is very large, this suggests that RM codes have better performance than (standard) polar codes under the ML decoder.
	
	The rest of this paper is organized as follows: In the next section, we introduce the notation and present necessary background on polar codes and RM codes. In Section~\ref{sect:mismatch}, we explain the mismatch between the standard polar code construction and the SCL decoder in detail. In Section~\ref{sect:DP}, we present our DP-polar code construction. Finally, the simulation results are given in Section~\ref{sect:simu}.

	\section{Background on polar codes and Reed-Muller codes} \label{sect:bg}
	Let $W:\{0,1\}\to\cY$ be a binary-input discrete memoryless channel with transition probabilities $\{W(y|x):x\in\{0,1\},y\in\cY\}$. We say that $W$ is a binary-input memoryless symmetric (BMS) channel if there is a permutation $\pi$ on the output alphabet $\cY$ such that i) $\pi^{-1}=\pi$ and ii) $W(y|1)=W(\pi(y)|0)$ for all $y\in\cY$.
	
	\begin{figure}[h!]
		\centering
		\begin{tikzpicture}
			\draw
			node at (0, 10.5) [] (u1)  {$U_0$}
			node at (0, 9) [] (u2)  {$U_1$}
			node at (1.5, 10.5) [XOR,scale=1.2] (x1) {}
			node at (2.5, 10.5) [] (xx1)  {$X_0$}
			node at (2.5, 9) [] (xx2)  {$X_1$}
			node at (3.8, 10.5) [block] (v1)  {$W$}
			node at (3.8, 9) [block] (v2)  {$W$}
			node at (5.2, 10.5) [] (y1)  {$Y_0$}
			node at (5.2, 9) [] (y2)  {$Y_1$};
			\draw[fill] (1.5, 9) circle (.5ex);
			\draw[very thick,->](u1) -- node {}(x1);
			\draw[very thick,->](u2) -| node {}(x1);
			\draw[very thick,->](x1) -- (xx1);
			\draw[very thick,->](u2) -- (xx2);
			\draw[very thick,->](xx1) -- (v1);
			\draw[very thick,->](xx2) -- (v2);
			\draw[very thick,->](v1) -- node {}(y1);
			\draw[very thick,->](v2) -- node {}(y2);
		\end{tikzpicture}
	\end{figure}
	
	The basic $2\times 2$ polar transform is illustrated in the figure above: Let $U_0,U_1$ be two i.i.d. Bernoulli-$1/2$ random variables. Let $X_0=U_0+U_1$ and $X_1=U_1$ be transmitted through two independent copies of $W$, and denote the channel outputs as $Y_0$ and $Y_1$, respectively. In this way, two copies of $W$ are transformed into two synthetic channels $W^-:U_0\to Y_0,Y_1$ and $W^+:U_1\to U_0,Y_0,Y_1$ defined as follows:
	\begin{equation} \label{eq:plms}
		\begin{aligned}
			& W^-(y_0,y_1|u_0)  = \frac{1}{2} \sum_{u_1\in\{0,1\}} W(y_0|u_0+u_1) W(y_1|u_1) \quad \text{for~} u_0\in\{0,1\} \text{~and~} y_0,y_1\in\cY , \\
			& W^+(u_0,y_0,y_1|u_1)  = \frac{1}{2} W(y_0|u_0+u_1) W(y_1|u_1) \quad \text{for~} u_0,u_1\in\{0,1\} \text{~and~} y_0,y_1\in\cY .
		\end{aligned}
	\end{equation}
	In order to obtain polar codes with code length $n=2^m$, we apply the $m$-step polar transform $\begin{bmatrix}
		1 & 0 \\
		1 & 1
	\end{bmatrix}^{\otimes m}$. More precisely, let $(U_0,U_1,\dots,U_{n-1})$ be the message vector and let
	$$
	(X_0,X_1,\dots,X_{n-1})=(U_0,U_1,\dots,U_{n-1})
	\begin{bmatrix}
		1 & 0 \\
		1 & 1
	\end{bmatrix}^{\otimes m} 
	$$
	be the codeword. Let $(Y_0,Y_1,\dots,Y_{n-1})$ be the channel output vector after transmitting $(X_0,X_1,\dots,X_{n-1})$ through $n$ independent copies of $W$. It can be shown that the set of synthetic channels $\{W_i:U_i\to U_0,\dots,U_{i-1},Y_0,\dots,Y_{n-1}\}_{i=0}^{n-1}$ has a one-to-one correspondence with $\{W^s:s\in\{+,-\}^m\}$, the set consisting of the $2^m$ synthetic channels obtained from $m$ steps of polar transforms to the channel $W$. When $m\to\infty$, it was shown in \cite{Arikan09} that almost all the $2^m$ synthetic channels are either noiseless or completely noisy. In the polar code construction, the bits corresponding to the noiseless synthetic channels are the information bits, and the ones corresponding to the completely noisy synthetic channels are the frozen bits. The information bits carry the message while the frozen bits are set to some fixed value, e.g., $0$.
	
	\begin{figure}[H]
		\centering
		{\footnotesize
			\begin{tikzpicture}[scale=0.75]
				\draw
				node at (-1.7, 12) [] ()  {\textcolor{blue}{Frozen}}
				node at (-1.7, 11.2) [] ()  {\textcolor{blue}{Frozen}}
				node at (-1.7, 10.4) [] () {\textcolor{blue}{Frozen}}
				node at (-1.7, 9.6) []  () {\textcolor{red}{Information}}
				node at (-1.7, 8.8) []  () {\textcolor{blue}{Frozen}}
				node at (-1.7, 8) []  () {\textcolor{red}{Information}}
				node at (-1.7, 7.2) [] ()  {\textcolor{red}{Information}}
				node at (-1.7, 6.4) [] ()  {\textcolor{red}{Information}}
				node at (-1.7, 5.6) [] ()  {\textcolor{blue}{Frozen}}
				node at (-1.7, 4.8) [] ()  {\textcolor{red}{Information}}
				node at (-1.7, 4) [] () {\textcolor{red}{Information}}
				node at (-1.7, 3.2) []  () {\textcolor{red}{Information}}
				node at (-1.7, 2.4) []  () {\textcolor{red}{Information}}
				node at (-1.7, 1.6) []  () {\textcolor{red}{Information}}
				node at (-1.7, 0.8) [] ()  {\textcolor{red}{Information}}
				node at (-1.7, 0) [] ()  {\textcolor{red}{Information}};
				
				\draw
				node at (0, 12) [] (u0)  {\textcolor{blue}{$U_0$}}
				node at (0, 11.2) [] (u1)  {\textcolor{blue}{$U_1$}}
				node at (0, 10.4) [] (u2) {\textcolor{blue}{$U_2$}}
				node at (0, 9.6) []  (u3) {\textcolor{red}{$U_3$}}
				node at (0, 8.8) []  (u4) {\textcolor{blue}{$U_4$}}
				node at (0, 8) []  (u5) {\textcolor{red}{$U_5$}}
				node at (0, 7.2) [] (u6)  {\textcolor{red}{$U_6$}}
				node at (0, 6.4) [] (u7)  {\textcolor{red}{$U_7$}}
				node at (0, 5.6) [] (u8)  {\textcolor{blue}{$U_8$}}
				node at (0, 4.8) [] (u9)  {\textcolor{red}{$U_9$}}
				node at (0, 4) [] (u10) {\textcolor{red}{$U_{10}$}}
				node at (0, 3.2) []  (u11) {\textcolor{red}{$U_{11}$}}
				node at (0, 2.4) []  (u12) {\textcolor{red}{$U_{12}$}}
				node at (0, 1.6) []  (u13) {\textcolor{red}{$U_{13}$}}
				node at (0, 0.8) [] (u14)  {\textcolor{red}{$U_{14}$}}
				node at (0, 0) [] (u15)  {\textcolor{red}{$U_{15}$}}
				
				node at (1, 12) [XOR, scale=0.75] (x0) {}
				node at (1, 10.4) [XOR, scale=0.75] (x2) {}
				node at (1, 8.8) [XOR, scale=0.75] (x4) {}
				node at (1, 7.2) [XOR, scale=0.75] (x6)  {}
				node at (1, 5.6) [XOR, scale=0.75] (x8) {}
				node at (1, 4) [XOR, scale=0.75] (x10) {}
				node at (1, 2.4) [XOR, scale=0.75] (x12) {}
				node at (1, 0.8) [XOR, scale=0.75] (x14)  {}
				
				node at (2, 12) [] (v0)  {$X_0^{(1)}$}
				node at (2, 11.2) [] (v1)  {$X_1^{(1)}$}
				node at (2, 10.4) [] (v2) {$X_2^{(1)}$}
				node at (2, 9.6) []  (v3) {$X_3^{(1)}$}
				node at (2, 8.8) []  (v4) {$X_4^{(1)}$}
				node at (2, 8) []  (v5) {$X_5^{(1)}$}
				node at (2, 7.2) [] (v6)  {$X_6^{(1)}$}
				node at (2, 6.4) [] (v7)  {$X_7^{(1)}$}
				node at (2, 5.6) [] (v8)  {$X_8^{(1)}$}
				node at (2, 4.8) [] (v9)  {$X_9^{(1)}$}
				node at (2, 4) [] (v10) {$X_{10}^{(1)}$}
				node at (2, 3.2) []  (v11) {$X_{11}^{(1)}$}
				node at (2, 2.4) []  (v12) {$X_{12}^{(1)}$}
				node at (2, 1.6) []  (v13) {$X_{13}^{(1)}$}
				node at (2, 0.8) [] (v14)  {$X_{14}^{(1)}$}
				node at (2, 0) [] (v15)  {$X_{15}^{(1)}$};
				
				\draw[thick,->](u0) -- node {}(x0);
				\draw[thick,->](u1) -| node {}(x0);
				\draw[thick,->](x0) -- node {}(v0);
				\draw[thick,->](u1) -- node {}(v1);
				\draw[fill] (1, 11.2) circle (.5ex);
				
				\draw[thick,->](u2) -- node {}(x2);
				\draw[thick,->](u3) -| node {}(x2);
				\draw[thick,->](x2) -- node {}(v2);
				\draw[thick,->](u3) -- node {}(v3);
				\draw[fill] (1, 9.6) circle (.5ex);
				
				\draw[thick,->](u4) -- node {}(x4);
				\draw[thick,->](u5) -| node {}(x4);
				\draw[thick,->](x4) -- node {}(v4);
				\draw[thick,->](u5) -- node {}(v5);
				\draw[fill] (1, 8) circle (.5ex);
				
				\draw[thick,->](u6) -- node {}(x6);
				\draw[thick,->](u7) -| node {}(x6);
				\draw[thick,->](x6) -- node {}(v6);
				\draw[thick,->](u7) -- node {}(v7);
				\draw[fill] (1, 6.4) circle (.5ex);
				
				\draw[thick,->](u8) -- node {}(x8);
				\draw[thick,->](u9) -| node {}(x8);
				\draw[thick,->](x8) -- node {}(v8);
				\draw[thick,->](u9) -- node {}(v9);
				\draw[fill] (1, 4.8) circle (.5ex);
				
				\draw[thick,->](u10) -- node {}(x10);
				\draw[thick,->](u11) -| node {}(x10);
				\draw[thick,->](x10) -- node {}(v10);
				\draw[thick,->](u11) -- node {}(v11);
				\draw[fill] (1, 3.2) circle (.5ex);
				
				\draw[thick,->](u12) -- node {}(x12);
				\draw[thick,->](u13) -| node {}(x12);
				\draw[thick,->](x12) -- node {}(v12);
				\draw[thick,->](u13) -- node {}(v13);
				\draw[fill] (1, 1.6) circle (.5ex);
				
				\draw[thick,->](u14) -- node {}(x14);
				\draw[thick,->](u15) -| node {}(x14);
				\draw[thick,->](x14) -- node {}(v14);
				\draw[thick,->](u15) -- node {}(v15);
				\draw[fill] (1, 0) circle (.5ex);

				\draw
				node at (3, 12) [XOR, scale=0.75] (xx0) {}
				node at (3, 8.8) [XOR, scale=0.75] (xx4) {}
				node at (3, 5.6) [XOR, scale=0.75] (xx8) {}
				node at (3, 2.4) [XOR, scale=0.75] (xx12)  {}
				node at (3.5, 11.2) [XOR, scale=0.75] (xx1) {}
				node at (3.5, 8) [XOR, scale=0.75] (xx5) {}
				node at (3.5, 4.8) [XOR, scale=0.75] (xx9) {}
				node at (3.5, 1.6) [XOR, scale=0.75] (xx13)  {}
				
				node at (4.5, 12) [] (vv0)  {$X_0^{(2)}$}
				node at (4.5, 11.2) [] (vv1)  {$X_1^{(2)}$}
				node at (4.5, 10.4) [] (vv2) {$X_2^{(2)}$}
				node at (4.5, 9.6) []  (vv3) {$X_3^{(2)}$}
				node at (4.5, 8.8) []  (vv4) {$X_4^{(2)}$}
				node at (4.5, 8) []  (vv5) {$X_5^{(2)}$}
				node at (4.5, 7.2) [] (vv6)  {$X_6^{(2)}$}
				node at (4.5, 6.4) [] (vv7)  {$X_7^{(2)}$}
				node at (4.5, 5.6) [] (vv8)  {$X_8^{(2)}$}
				node at (4.5, 4.8) [] (vv9)  {$X_9^{(2)}$}
				node at (4.5, 4) [] (vv10) {$X_{10}^{(2)}$}
				node at (4.5, 3.2) []  (vv11) {$X_{11}^{(2)}$}
				node at (4.5, 2.4) []  (vv12) {$X_{12}^{(2)}$}
				node at (4.5, 1.6) []  (vv13) {$X_{13}^{(2)}$}
				node at (4.5, 0.8) [] (vv14)  {$X_{14}^{(2)}$}
				node at (4.5, 0) [] (vv15)  {$X_{15}^{(2)}$};
				
				\draw[thick,->](v0) -- node {}(xx0);
				\draw[thick,->](v2) -| node {}(xx0);
				\draw[thick,->](xx0) -- node {}(vv0);
				\draw[thick,->](v2) -- node {}(vv2);
				\draw[thick,->](v1) -- node {}(xx1);
				\draw[thick,->](v3) -| node {}(xx1);
				\draw[thick,->](xx1) -- node {}(vv1);
				\draw[thick,->](v3) -- node {}(vv3);
				\draw[fill] (3, 10.4) circle (.5ex);
				\draw[fill] (3.5, 9.6) circle (.5ex);
				
				\draw[thick,->](v4) -- node {}(xx4);
				\draw[thick,->](v6) -| node {}(xx4);
				\draw[thick,->](xx4) -- node {}(vv4);
				\draw[thick,->](v6) -- node {}(vv6);
				\draw[thick,->](v5) -- node {}(xx5);
				\draw[thick,->](v7) -| node {}(xx5);
				\draw[thick,->](xx5) -- node {}(vv5);
				\draw[thick,->](v7) -- node {}(vv7);
				\draw[fill] (3, 7.2) circle (.5ex);
				\draw[fill] (3.5, 6.4) circle (.5ex);
				
				\draw[thick,->](v8) -- node {}(xx8);
				\draw[thick,->](v10) -| node {}(xx8);
				\draw[thick,->](xx8) -- node {}(vv8);
				\draw[thick,->](v10) -- node {}(vv10);
				\draw[thick,->](v9) -- node {}(xx9);
				\draw[thick,->](v11) -| node {}(xx9);
				\draw[thick,->](xx9) -- node {}(vv9);
				\draw[thick,->](v11) -- node {}(vv11);
				\draw[fill] (3, 4) circle (.5ex);
				\draw[fill] (3.5, 3.2) circle (.5ex);
				
				\draw[thick,->](v12) -- node {}(xx12);
				\draw[thick,->](v14) -| node {}(xx12);
				\draw[thick,->](xx12) -- node {}(vv12);
				\draw[thick,->](v14) -- node {}(vv14);
				\draw[thick,->](v13) -- node {}(xx13);
				\draw[thick,->](v15) -| node {}(xx13);
				\draw[thick,->](xx13) -- node {}(vv13);
				\draw[thick,->](v15) -- node {}(vv15);
				\draw[fill] (3, 0.8) circle (.5ex);
				\draw[fill] (3.5, 0) circle (.5ex);

				\draw
				node at (5.5, 12) [XOR, scale=0.75] (xxx0) {}
				node at (5.5, 5.6) [XOR, scale=0.75] (xxx8) {}
				node at (6, 11.2) [XOR, scale=0.75] (xxx1) {}
				node at (6, 4.8) [XOR, scale=0.75] (xxx9)  {}
				node at (6.5, 10.4) [XOR, scale=0.75] (xxx2) {}
				node at (6.5, 4) [XOR, scale=0.75] (xxx10) {}
				node at (7, 9.6) [XOR, scale=0.75] (xxx3) {}
				node at (7, 3.2) [XOR, scale=0.75] (xxx11)  {}
				
				node at (8, 12) [] (vvv0)  {$X_0^{(3)}$}
				node at (8, 11.2) [] (vvv1)  {$X_1^{(3)}$}
				node at (8, 10.4) [] (vvv2) {$X_2^{(3)}$}
				node at (8, 9.6) []  (vvv3) {$X_3^{(3)}$}
				node at (8, 8.8) []  (vvv4) {$X_4^{(3)}$}
				node at (8, 8) []  (vvv5) {$X_5^{(3)}$}
				node at (8, 7.2) [] (vvv6)  {$X_6^{(3)}$}
				node at (8, 6.4) [] (vvv7)  {$X_7^{(3)}$}
				node at (8, 5.6) [] (vvv8)  {$X_8^{(3)}$}
				node at (8, 4.8) [] (vvv9)  {$X_9^{(3)}$}
				node at (8, 4) [] (vvv10) {$X_{10}^{(3)}$}
				node at (8, 3.2) []  (vvv11) {$X_{11}^{(3)}$}
				node at (8, 2.4) []  (vvv12) {$X_{12}^{(3)}$}
				node at (8, 1.6) []  (vvv13) {$X_{13}^{(3)}$}
				node at (8, 0.8) [] (vvv14)  {$X_{14}^{(3)}$}
				node at (8, 0) [] (vvv15)  {$X_{15}^{(3)}$};
				
				\draw[thick,->](vv0) -- node {}(xxx0);
				\draw[thick,->](vv4) -| node {}(xxx0);
				\draw[thick,->](xxx0) -- node {}(vvv0);
				\draw[thick,->](vv4) -- node {}(vvv4);
				\draw[fill] (5.5, 8.8) circle (.5ex);
				
				\draw[thick,->](vv1) -- node {}(xxx1);
				\draw[thick,->](vv5) -| node {}(xxx1);
				\draw[thick,->](xxx1) -- node {}(vvv1);
				\draw[thick,->](vv5) -- node {}(vvv5);
				\draw[fill] (6, 8) circle (.5ex);
				
				\draw[thick,->](vv2) -- node {}(xxx2);
				\draw[thick,->](vv6) -| node {}(xxx2);
				\draw[thick,->](xxx2) -- node {}(vvv2);
				\draw[thick,->](vv6) -- node {}(vvv6);
				\draw[fill] (6.5, 7.2) circle (.5ex);
				
				\draw[thick,->](vv3) -- node {}(xxx3);
				\draw[thick,->](vv7) -| node {}(xxx3);
				\draw[thick,->](xxx3) -- node {}(vvv3);
				\draw[thick,->](vv7) -- node {}(vvv7);
				\draw[fill] (7, 6.4) circle (.5ex);

				\draw[thick,->](vv8) -- node {}(xxx8);
				\draw[thick,->](vv12) -| node {}(xxx8);
				\draw[thick,->](xxx8) -- node {}(vvv8);
				\draw[thick,->](vv12) -- node {}(vvv12);
				\draw[fill] (5.5, 2.4) circle (.5ex);
				
				\draw[thick,->](vv9) -- node {}(xxx9);
				\draw[thick,->](vv13) -| node {}(xxx9);
				\draw[thick,->](xxx9) -- node {}(vvv9);
				\draw[thick,->](vv13) -- node {}(vvv13);
				\draw[fill] (6, 1.6) circle (.5ex);
				
				\draw[thick,->](vv10) -- node {}(xxx10);
				\draw[thick,->](vv14) -| node {}(xxx10);
				\draw[thick,->](xxx10) -- node {}(vvv10);
				\draw[thick,->](vv14) -- node {}(vvv14);
				\draw[fill] (6.5, 0.8) circle (.5ex);
				
				\draw[thick,->](vv11) -- node {}(xxx11);
				\draw[thick,->](vv15) -| node {}(xxx11);
				\draw[thick,->](xxx11) -- node {}(vvv11);
				\draw[thick,->](vv15) -- node {}(vvv15);
				\draw[fill] (7, 0) circle (.5ex);

				\draw
				node at (9, 12) [XOR, scale=0.75] (xxxx0) {}
				node at (9.5, 11.2) [XOR, scale=0.75] (xxxx1) {}
				node at (10, 10.4) [XOR, scale=0.75] (xxxx2) {}
				node at (10.5, 9.6) [XOR, scale=0.75] (xxxx3)  {}
				node at (11, 8.8) [XOR, scale=0.75] (xxxx4) {}
				node at (11.5, 8) [XOR, scale=0.75] (xxxx5) {}
				node at (12, 7.2) [XOR, scale=0.75] (xxxx6) {}
				node at (12.5, 6.4) [XOR, scale=0.75] (xxxx7) {}
				
				node at (13.3, 12) [] (vvvv0)  {$X_0$}
				node at (13.3, 11.2) [] (vvvv1)  {$X_1$}
				node at (13.3, 10.4) [] (vvvv2) {$X_2$}
				node at (13.3, 9.6) []  (vvvv3) {$X_3$}
				node at (13.3, 8.8) []  (vvvv4) {$X_4$}
				node at (13.3, 8) []  (vvvv5) {$X_5$}
				node at (13.3, 7.2) [] (vvvv6)  {$X_6$}
				node at (13.3, 6.4) [] (vvvv7)  {$X_7$}
				node at (13.3, 5.6) [] (vvvv8)  {$X_8$}
				node at (13.3, 4.8) [] (vvvv9)  {$X_9$}
				node at (13.3, 4) [] (vvvv10) {$X_{10}$}
				node at (13.3, 3.2) []  (vvvv11) {$X_{11}$}
				node at (13.3, 2.4) []  (vvvv12) {$X_{12}$}
				node at (13.3, 1.6) []  (vvvv13) {$X_{13}$}
				node at (13.3, 0.8) [] (vvvv14)  {$X_{14}$}
				node at (13.3, 0) [] (vvvv15)  {$X_{15}$};

				\draw[thick,->](vvv0) -- node {}(xxxx0);
				\draw[thick,->](vvv8) -| node {}(xxxx0);
				\draw[thick,->](xxxx0) -- node {}(vvvv0);
				\draw[thick,->](vvv8) -- node {}(vvvv8);
				\draw[fill] (9, 5.6) circle (.5ex);
				
				\draw[thick,->](vvv1) -- node {}(xxxx1);
				\draw[thick,->](vvv9) -| node {}(xxxx1);
				\draw[thick,->](xxxx1) -- node {}(vvvv1);
				\draw[thick,->](vvv9) -- node {}(vvvv9);
				\draw[fill] (9.5, 4.8) circle (.5ex);
				
				\draw[thick,->](vvv2) -- node {}(xxxx2);
				\draw[thick,->](vvv10) -| node {}(xxxx2);
				\draw[thick,->](xxxx2) -- node {}(vvvv2);
				\draw[thick,->](vvv10) -- node {}(vvvv10);
				\draw[fill] (10, 4) circle (.5ex);
				
				\draw[thick,->](vvv3) -- node {}(xxxx3);
				\draw[thick,->](vvv11) -| node {}(xxxx3);
				\draw[thick,->](xxxx3) -- node {}(vvvv3);
				\draw[thick,->](vvv11) -- node {}(vvvv11);
				\draw[fill] (10.5, 3.2) circle (.5ex);
				
				\draw[thick,->](vvv4) -- node {}(xxxx4);
				\draw[thick,->](vvv12) -| node {}(xxxx4);
				\draw[thick,->](xxxx4) -- node {}(vvvv4);
				\draw[thick,->](vvv12) -- node {}(vvvv12);
				\draw[fill] (11, 2.4) circle (.5ex);
				
				\draw[thick,->](vvv5) -- node {}(xxxx5);
				\draw[thick,->](vvv13) -| node {}(xxxx5);
				\draw[thick,->](xxxx5) -- node {}(vvvv5);
				\draw[thick,->](vvv13) -- node {}(vvvv13);
				\draw[fill] (11.5, 1.6) circle (.5ex);
				
				\draw[thick,->](vvv6) -- node {}(xxxx6);
				\draw[thick,->](vvv14) -| node {}(xxxx6);
				\draw[thick,->](xxxx6) -- node {}(vvvv6);
				\draw[thick,->](vvv14) -- node {}(vvvv14);
				\draw[fill] (12, 0.8) circle (.5ex);
				
				\draw[thick,->](vvv7) -- node {}(xxxx7);
				\draw[thick,->](vvv15) -| node {}(xxxx7);
				\draw[thick,->](xxxx7) -- node {}(vvvv7);
				\draw[thick,->](vvv15) -- node {}(vvvv15);
				\draw[fill] (12.5, 0) circle (.5ex);

				\draw
				node at (14.5, 12) [block, scale=0.75] (w0)  {$W$}
				node at (14.5, 11.2) [block, scale=0.75] (w1)  {$W$}
				node at (14.5, 10.4) [block, scale=0.75] (w2) {$W$}
				node at (14.5, 9.6) [block, scale=0.75]  (w3) {$W$}
				node at (14.5, 8.8) [block, scale=0.75]  (w4) {$W$}
				node at (14.5, 8) [block, scale=0.75]  (w5) {$W$}
				node at (14.5, 7.2) [block, scale=0.75] (w6)  {$W$}
				node at (14.5, 6.4) [block, scale=0.75] (w7)  {$W$}
				node at (14.5, 5.6) [block, scale=0.75] (w8)  {$W$}
				node at (14.5, 4.8) [block, scale=0.75] (w9)  {$W$}
				node at (14.5, 4) [block, scale=0.75] (w10) {$W$}
				node at (14.5, 3.2) [block, scale=0.75]  (w11) {$W$}
				node at (14.5, 2.4) [block, scale=0.75]  (w12) {$W$}
				node at (14.5, 1.6) [block, scale=0.75]  (w13) {$W$}
				node at (14.5, 0.8) [block, scale=0.75] (w14)  {$W$}
				node at (14.5, 0) [block, scale=0.75] (w15)  {$W$}
				
				node at (15.7, 12) [] (y0)  {$Y_0$}
				node at (15.7, 11.2) [] (y1)  {$Y_1$}
				node at (15.7, 10.4) [] (y2) {$Y_2$}
				node at (15.7, 9.6) []  (y3) {$Y_3$}
				node at (15.7, 8.8) []  (y4) {$Y_4$}
				node at (15.7, 8) []  (y5) {$Y_5$}
				node at (15.7, 7.2) [] (y6)  {$Y_6$}
				node at (15.7, 6.4) [] (y7)  {$Y_7$}
				node at (15.7, 5.6) [] (y8)  {$Y_8$}
				node at (15.7, 4.8) [] (y9)  {$Y_9$}
				node at (15.7, 4) [] (y10) {$Y_{10}$}
				node at (15.7, 3.2) []  (y11) {$Y_{11}$}
				node at (15.7, 2.4) []  (y12) {$Y_{12}$}
				node at (15.7, 1.6) []  (y13) {$Y_{13}$}
				node at (15.7, 0.8) [] (y14)  {$Y_{14}$}
				node at (15.7, 0) [] (y15)  {$Y_{15}$};

				\draw[thick,->](vvvv0) -- node {}(w0);
				\draw[thick,->](vvvv1) -- node {}(w1);
				\draw[thick,->](vvvv2) -- node {}(w2);
				\draw[thick,->](vvvv3) -- node {}(w3);
				\draw[thick,->](vvvv4) -- node {}(w4);
				\draw[thick,->](vvvv5) -- node {}(w5);
				\draw[thick,->](vvvv6) -- node {}(w6);
				\draw[thick,->](vvvv7) -- node {}(w7);
				\draw[thick,->](vvvv8) -- node {}(w8);
				\draw[thick,->](vvvv9) -- node {}(w9);
				\draw[thick,->](vvvv10) -- node {}(w10);
				\draw[thick,->](vvvv11) -- node {}(w11);
				\draw[thick,->](vvvv12) -- node {}(w12);
				\draw[thick,->](vvvv13) -- node {}(w13);
				\draw[thick,->](vvvv14) -- node {}(w14);
				\draw[thick,->](vvvv15) -- node {}(w15);

				\draw[thick,->](w0) -- node {}(y0);
				\draw[thick,->](w1) -- node {}(y1);
				\draw[thick,->](w2) -- node {}(y2);
				\draw[thick,->](w3) -- node {}(y3);
				\draw[thick,->](w4) -- node {}(y4);
				\draw[thick,->](w5) -- node {}(y5);
				\draw[thick,->](w6) -- node {}(y6);
				\draw[thick,->](w7) -- node {}(y7);
				\draw[thick,->](w8) -- node {}(y8);
				\draw[thick,->](w9) -- node {}(y9);
				\draw[thick,->](w10) -- node {}(y10);
				\draw[thick,->](w11) -- node {}(y11);
				\draw[thick,->](w12) -- node {}(y12);
				\draw[thick,->](w13) -- node {}(y13);
				\draw[thick,->](w14) -- node {}(y14);
				\draw[thick,->](w15) -- node {}(y15);

				\draw
				node at (-1.1, -1.2) [] (p0)  {\textcolor{blue}{Polar$(1,0)$}}
				node at (-1.1, -2) [] (p1)  {\textcolor{blue}{Polar$(1,0)$}}
				node at (-1.1, -2.8) [] (p2)  {\textcolor{blue}{Polar$(1,0)$}}
				node at (-1.1, -3.6) [] (p3)  {\textcolor{red}{Polar$(1,1)$}}
				node at (-1.1, -4.4) [] (p4)  {\textcolor{blue}{Polar$(1,0)$}}
				node at (-1.1, -5.2) [] (p5)  {\textcolor{red}{Polar$(1,1)$}}
				node at (-1.1, -6) [] (p6)  {\textcolor{red}{Polar$(1,1)$}}
				node at (-1.1, -6.8) [] (p7)  {\textcolor{red}{Polar$(1,1)$}}
				node at (-1.1, -7.6) [] (p8)  {\textcolor{blue}{Polar$(1,0)$}}
				node at (-1.1, -8.4) [] (p9)  {\textcolor{red}{Polar$(1,1)$}}
				node at (-1.1, -9.2) [] (p10)  {\textcolor{red}{Polar$(1,1)$}}
				node at (-1.1, -10) [] (p11)  {\textcolor{red}{Polar$(1,1)$}}
				node at (-1.1, -10.8) [] (p12)  {\textcolor{red}{Polar$(1,1)$}}
				node at (-1.1, -11.6) [] (p13)  {\textcolor{red}{Polar$(1,1)$}}
				node at (-1.1, -12.4) [] (p14)  {\textcolor{red}{Polar$(1,1)$}}
				node at (-1.1, -13.2) [] (p15)  {\textcolor{red}{Polar$(1,1)$}}
				
				node at (1.7, -1.6) [] (pp0) {Polar$(2,0)$}
				node at (1.7, -3.2) [] (pp2) {Polar$(2,1)$}
				node at (1.7, -4.8) [] (pp4) {Polar$(2,1)$}
				node at (1.7, -6.4) [] (pp6) {Polar$(2,2)$}
				node at (1.7, -8) [] (pp8) {Polar$(2,1)$}
				node at (1.7, -9.6) [] (pp10) {Polar$(2,2)$}
				node at (1.7, -11.2) [] (pp12){Polar$(2,2)$}
				node at (1.7, -12.8) [] (pp14){Polar$(2,2)$}
				
				node at (4.5, -2.4) [] (t0){Polar$(4,1)$}
				node at (4.5, -5.6) [] (t4){Polar$(4,3)$}
				node at (4.5, -8.8) [] (t8){Polar$(4,3)$}
				node at (4.5, -12) [] (t12){Polar$(4,4)$}
				
				node at (8, -4) [] (tt0){Polar$(8,4)$}
				node at (8, -10.4) [] (tt8){Polar$(8,7)$}
				
				node at (13.3, -7.2) [] (ttt0) {Polar$(16,11)$};

				\draw (p0) -- node {}(pp0);
				\draw (p1) -- node {}(pp0);
				\draw (p2) -- node {}(pp2);
				\draw (p3) -- node {}(pp2);
				\draw (p4) -- node {}(pp4);
				\draw (p5) -- node {}(pp4);
				\draw (p6) -- node {}(pp6);
				\draw (p7) -- node {}(pp6);
				\draw (p8) -- node {}(pp8);
				\draw (p9) -- node {}(pp8);
				\draw (p10) -- node {}(pp10);
				\draw (p11) -- node {}(pp10);
				\draw (p12) -- node {}(pp12);
				\draw (p13) -- node {}(pp12);
				\draw (p14) -- node {}(pp14);
				\draw (p15) -- node {}(pp14);
				
				\draw (pp0) -- node {}(t0);
				\draw (pp2) -- node {}(t0);
				\draw (pp4) -- node {}(t4);
				\draw (pp6) -- node {}(t4);
				\draw (pp8) -- node {}(t8);
				\draw (pp10) -- node {}(t8);
				\draw (pp12) -- node {}(t12);
				\draw (pp14) -- node {}(t12);
				
				\draw (t0) -- node {}(tt0);
				\draw (t4) -- node {}(tt0);
				\draw (t8) -- node {}(tt8);
				\draw (t12) -- node {}(tt8);
				
				\draw (tt0) -- node {}(ttt0);
				\draw (tt8) -- node {}(ttt0);

			\end{tikzpicture}
		}
		\caption{The $(16,11)$ polar code constructed for the Binary-Input AWGN channel $W$ with $E_b/N_0=2\dB$. Each blue $U_i$ is a frozen bit, and it can be viewed as a $(1,0)$ polar code; each red $U_i$ is an information bit, and it can be viewed as a $(1,1)$ polar code. In the first layer of polar transforms, $16$ length-$1$ polar codes are divided into $8$ pairs, and each pair is transformed into a length-$2$ polar code. Then in the second layer, eight length-$2$ polar codes are transformed into four polar codes with length $4$. This process continues until we eventually combine two length-$8$ polar codes to obtain the $(16,11)$ polar code.}
		\label{fig:fceng}
	\end{figure}

	Fig.~\ref{fig:fceng} shows a concrete example of polar code construction for $m=4$. The $16$ synthetic channels in this figure are $W_i:U_i\to U_0,\dots,U_{i-1},Y_0,\dots,Y_{15}$ for $0\le i\le 15$. For every $0\le i\le 15$, let $s(i)\in\{+,-\}^4$ be the sequence obtained by replacing ``$0$" with ``$-$" and replacing ``$1$" with ``$+$" in the $4$-digit binary expansion of $i$. For example, $s(1)$ is $---+$, and $s(6)$ is $-++-$. The synthetic channel $W_i$ is the same as $W^{s(i)}$ for every $0\le i\le 15$, where $W^{s(i)}$ is obtained by recursively applying the definition in \eqref{eq:plms}. This relation allows us to calculate a tight approximation of the channel capacity of each synthetic channel $W_i$ using the method in \cite{Tal13}. The capacity of each $W_i$ is then used to decide whether $U_i$ is an information bit or a frozen bit. More precisely, let $\cA$ be the index set of the information bits, i.e., $\{U_i:i\in\cA\}$ are the information bits, and $|\cA|$ is equal to the code dimension $k$. In the standard construction of an $(n=2^m,k)$ polar code for a given BMS channel $W$, the set $\cA$ is chosen to satisfy that the $k$ synthetic channels $\{W_i:i\in\cA\}$ have the largest channel capacity among all the $n$ synthetic channels.  
	
	The example in Fig.~\ref{fig:fceng} is an $(n=16,k=11)$ polar code constructed for the Binary-Input AWGN channel $W$ with $E_b/N_0=2\dB$. In this example, $\cA=\{3,5,6,7,9,10,11,12,13,14,15\}$. On the left side of the figure, each frozen bit can be viewed as a $(1,0)$ polar code, i.e., a polar code with length $1$ and dimension $0$. Similarly, each information bit can be viewed as a $(1,1)$ polar code. In the first layer of polar transforms, two frozen bits are transformed into a $(2,0)$ polar code, e.g., $(U_0,U_1)$ is transformed into $(X_0^{(1)},X_1^{(1)})$, which is a $(2,0)$ polar code; one frozen bit and one information bit are transformed into a $(2,1)$ polar code, e.g., $(U_2,U_3)$ is transformed into $(X_2^{(1)},X_3^{(1)})$; two information bits are transformed into a $(2,2)$ polar code, e.g., $(U_6,U_7)$ is transformed into $(X_6^{(1)},X_7^{(1)})$. In the second layer of polar transforms, two length-$2$ polar codes are transformed into a length-$4$ polar code. For example, the $(2,1)$ polar code $(X_4^{(1)},X_5^{(1)})$ and the $(2,2)$ polar code $(X_6^{(1)},X_7^{(1)})$ are transformed into the $(4,3)$ polar code $(X_4^{(2)},X_5^{(2)},X_6^{(2)},X_7^{(2)})$. In the next layer, two length-$4$ polar codes are further transformed into a length-$8$ polar codes, e.g., the $(4,1)$ polar code $(X_0^{(2)},X_1^{(2)},X_2^{(2)},X_3^{(2)})$ and the $(4,3)$ polar code $(X_4^{(2)},X_5^{(2)},X_6^{(2)},X_7^{(2)})$ are transformed into the $(8,4)$ polar code $(X_0^{(3)},X_1^{(3)},\dots,X_7^{(3)})$. Finally, in the last layer, the $(8,4)$ polar code $(X_0^{(3)},X_1^{(3)},\dots,X_7^{(3)})$ and the $(8,7)$ polar code $(X_8^{(3)},X_9^{(3)},\dots,X_{15}^{(3)})$ are transformed into the $(16,11)$ polar code $(X_0,X_1,\dots,X_{15})$.
	
	\subsection{Successive Cancellation decoder of polar codes}
	The Successive Cancellation (SC) decoder of polar codes works in a recursive way. When decoding a length-$n$ polar code, the SC decoder decomposes it into two length-$n/2$ polar codes and decodes these two codes one by one. For example, the $(16,11)$ polar code in Fig.~\ref{fig:fceng} can be decomposed into the $(8,4)$ polar code $(X_0^{(3)},X_1^{(3)},\dots,X_7^{(3)})$ and the $(8,7)$ polar code $(X_8^{(3)},X_9^{(3)},\dots,X_{15}^{(3)})$. The SC decoder first uses the channel output $(Y_0,Y_1,\dots,Y_{15})$ to decode the $(8,4)$ polar code $(X_0^{(3)},X_1^{(3)},\dots,X_7^{(3)})$. Let us denote the decoding result as $(\hat{X}_0^{(3)},\hat{X}_1^{(3)},\dots,\hat{X}_7^{(3)})$. Then the SC decoder uses both the channel output $(Y_0,Y_1,\dots,Y_{15})$ and the previous decoding result $(\hat{X}_0^{(3)},\hat{X}_1^{(3)},\dots,\hat{X}_7^{(3)})$ to decode the $(8,7)$ polar code $(X_8^{(3)},X_9^{(3)},\dots,X_{15}^{(3)})$, whose decoding result is denoted as $(\hat{X}_8^{(3)},\hat{X}_9^{(3)},\dots,\hat{X}_{15}^{(3)})$. Finally, the SC decoder applies the last layer of polar transform to $(\hat{X}_0^{(3)},\hat{X}_1^{(3)},\dots,\hat{X}_7^{(3)})$ and $(\hat{X}_8^{(3)},\hat{X}_9^{(3)},\dots,\hat{X}_{15}^{(3)})$ to obtain the final decoding result $(\hat{X}_0,\hat{X}_1,\dots,\hat{X}_{15})$. This describes the outer layer of the SC decoder. In the next layer, when decoding the two length-$8$ polar codes, the SC decoder follows the same recursive structure. More precisely, when decoding the $(8,4)$ polar code $(X_0^{(3)},X_1^{(3)},\dots,X_7^{(3)})$, the SC decoder first decodes the $(4,1)$ polar code $(X_0^{(2)},X_1^{(2)},X_2^{(2)},X_3^{(2)})$ and then decodes the $(4,3)$ polar code $(X_4^{(2)},X_5^{(2)},X_6^{(2)},X_7^{(2)})$. Similarly, when decoding the $(8,7)$ polar code $(X_8^{(3)},X_9^{(3)},\dots,X_{15}^{(3)})$, the SC decoder first decodes the $(4,3)$ polar code $(X_8^{(2)},X_9^{(2)},X_{10}^{(2)},X_{11}^{(2)})$ and then decodes the $(4,4)$ polar code $(X_{12}^{(2)},X_{13}^{(2)},X_{14}^{(2)},X_{15}^{(2)})$. This recursive procedure continues until the SC decoder reaches the length-$1$ polar codes, i.e., the frozen bits and the information bits $U_0,U_1,\dots,U_{15}$. To summarize, in the tree diagram in Fig.~\ref{fig:fceng}, the decoding procedure of the SC decoder goes from right to left and from top to bottom.
	
	For $0\le i\le 7$, the channel mapping from $X_i^{(3)}$ to $(Y_0,Y_1,\dots,Y_{15})$ is $W^-$ because only $Y_i$ and $Y_{i+8}$ depend on the value of $X_i^{(3)}$. For $8\le i\le 15$, the channel mapping from $X_i^{(3)}$ to $(X_0^{(3)},X_1^{(3)},\dots,X_7^{(3)},Y_0,\linebreak[4]
	Y_1,\dots,Y_{15})$ is $W^+$. Therefore, the $(8,4)$ polar code $(X_0^{(3)},X_1^{(3)},\dots,X_7^{(3)})$ in the top branch is constructed for $W^-$, and the $(8,7)$ polar code $(X_8^{(3)},X_9^{(3)},\dots,X_{15}^{(3)})$ in the bottom branch is constructed for $W^+$. In fact, every top branch in the tree diagram corresponds to the ``minus" transform on the channel, and every bottom branch corresponds to the ``plus" transform. For example, the $(4,1)$ polar code $(X_0^{(2)},X_1^{(2)},X_2^{(2)},X_3^{(2)})$ is constructed for $W^{--}$, and the $(2,1)$ polar code $(X_2^{(1)},X_3^{(1)})$ is constructed for $W^{--+}$. In this paper, we will use the terms ``top branch" and ``minus branch" interchangeably, and we also use the terms ``bottom branch" and ``plus branch" interchangeably.
	
	As a final remark, we note that the binary tree representation in Fig.~\ref{fig:fceng} has a natural connection with the Plotkin $(\mathbi{u},\mathbi{u}+\mathbi{v})$ decomposition: The top branch in the binary tree corresponds to the component $\mathbi{v}$, and the bottom branch corresponds to the component $\mathbi{u}$.
	
	\subsection{Reed-Muller codes} \label{sect:RM}
	The generator matrices of both polar codes and RM codes are submatrices of $G_m:=\begin{bmatrix}
		1 & 0 \\
		1 & 1
	\end{bmatrix}^{\otimes m}$. Polar codes select row vectors corresponding to noiseless synthetic channels while RM codes select row vectors with the largest Hamming weight. More precisely, we label the row indices of $G_m$ from $0$ to $2^m-1$. For $0\le i\le 2^m-1$, we use $\wt(i)$ to denote the Hamming weight of the $m$-digit binary expansion of $i$. Then the Hamming weight of the $i$th row of $G_m$ is $2^{\wt(i)}$. The generator matrix of the RM code with parameters $m$ and $r$ consists of all the row vectors of $G_m$ whose indices satisfy $\wt(i)\ge m-r$.
	
	Note that originally RM codes were defined using multivariate polynomials with binary coefficients \cite{Reed54,Muller54}, but for the purpose of this paper we will stick to the definition described above. Readers may consult a recent survey \cite{Abbe21} on RM codes to see why these two definitions are equivalent.

	\section{Mismatch between polar code construction and the SCL decoder} \label{sect:mismatch}
	
	Following the notation in Fig.~\ref{fig:fceng}, for a length-$n$ polar code, we use $(U_0,U_1,\dots,U_{n-1})$ to denote the vector containing the information bits and the frozen bits. We use $(X_0,X_1,\dots,X_{n-1})$ and $(Y_0,Y_1,\dots,Y_{n-1})$ to denote the codeword and the channel output vector, respectively. Given a BMS channel $W$, the code length $n$ and the code dimension $k$, polar code construction amounts to finding which $k$ bits in $(U_0,U_1,\dots,U_{n-1})$ are the information bits. In the standard polar code construction, e.g., the construction described in \cite{Tal13}, we calculate a tight approximation of the conditional entropy\footnote{Conditional entropy is not the only choice. We can also use other metrics to measure the uncertainty of $U_i$ given $U_0,U_1,\dots,U_{i-1},Y_0,Y_1,\dots,Y_{n-1}$, e.g., the Bhattacharyya parameter.}
	\begin{equation} \label{eq:hi}
		H_i := H(U_i|U_0,U_1,\dots,U_{i-1},Y_0,Y_1,\dots,Y_{n-1}) 
	\end{equation}
	and then choose the $k$ information bits as the ones corresponding to the smallest values of $H_i$'s. This code construction works perfectly for the SC decoder because the SC decoder only makes use of the previous message bits, i.e., $U_0,U_1,\dots,U_{i-1}$. However, there is a mismatch between this code construction and the SCL decoder because the list decoder also makes use of the information from the future frozen bits. To see this, let us first recall how the SCL decoder works. Similarly to the SC decoder, the SCL decoder also decodes from $U_0$ to $U_{n-1}$ one by one. The difference is that the SC decoder only keeps one candidate of the decoding results of previous bits while the SCL decoder keeps a list of multiple candidates. When reaching an information bit, the SCL decoder expand the current list size by a factor of $2$ to append both possible values (i.e., $0$ and $1$) of this information bit at the end of each candidate. A pruning procedure according to the likelihood of the candidates will be activated if the current list size becomes too large after the expansion. When reaching a frozen bit, the likelihood of each candidate in the list is updated based on the conditional probability of that frozen bit given the values of the previous bits and the channel outputs. This can penalize some candidates by a considerable amount if the
	values of the previously decoded bits in the candidate do not agree with the value of the frozen bit. In this way, the frozen bit serves as a ``soft parity check" for the previously decoded information bits, or in other words, the decoding decision on each information bit relies not only on the previous bits but also on the future frozen bits.
	
	Let us take the following extreme case as an example. When the list size is larger than $2^k$, the SCL decoder is the same as the Maximum Likelihood (ML) decoder. In this case, the uncertainty of each information bit under the SCL decoder is
	\begin{equation} \label{eq:hml}
		H_i^{\ML} := H(U_i|\{U_j:j\in\cA^c \},Y_0,Y_1,\dots,Y_{n-1}) ,
	\end{equation}
	where $\cA$ is the index set of information bits and $\cA^c$ is the index set of frozen bits. Notice that for the standard polar code construction,
	\begin{equation} \label{eq:hap}
		H_i \approx H(U_i|\{U_j:j\in\cA^c, j<i \},Y_0,Y_1,\dots,Y_{n-1}) ,
	\end{equation}
	where $H_i$ is defined in \eqref{eq:hi}. This is because for the standard polar code construction, the information bits $\{U_j:j\in\cA, j<i \}$ can be determined by the frozen bits $\{U_j:j\in\cA^c, j<i \}$ and the channel outputs $Y_0,Y_1,\dots,Y_{n-1}$ with high probability. Comparing \eqref{eq:hml} with \eqref{eq:hap}, we can see that $H_i^{\ML}$ is typically smaller than $H_i$, indicating that there is a mismatch between the polar code construction and the ML decoder.
	
	As for the normal SCL decoder whose list size is not as large as $2^k$, the uncertainty of each information bit under the SCL decoder can be approximated by the following conditional entropy
	\begin{equation} \label{eq:hscl}
		H_i^{\SCL} \approx H(U_i|\{U_j:j\in\cA^c, j<i+\Delta_{i,L} \},Y_0,Y_1,\dots,Y_{n-1}) ,
	\end{equation}
	where the nonnegative integer $\Delta_{i,L}$ depends on both the index $i$ and the list size $L$. By the discussion above, when $L\ge 2^k$, we can take $\Delta_{i,L}=n-i$. For this choice of $\Delta_{i,L}$, we have $\{U_j:j\in\cA^c, j<i+\Delta_{i,L} \}=\{U_j:j\in\cA^c \}$ and $H_i^{\SCL}=H_i^{\ML}$. When $L$ takes a smaller value, the dependence of $\Delta_{i,L}$ on $L$ is rather complicated to quantify. However, one would expect that $\Delta_{i,L}$ is an increasing function of $L$ because intuitively, a larger list size allows the SCL decoder to make use of information from more future frozen bits. In the extreme case of $L=1$, we have $\Delta_{i,L}=0$ and $H_i^{\SCL}\approx H_i$ because in this case the SCL decoder degenerates into the SC decoder.
	Now we conclude that
	$$
	H_i \ge H_i^{\SCL} \ge H_i^{\ML} .
	$$
	Moreover, when the list size is small, $H_i^{\SCL}$ tends to be close to $H_i$; when the list size is large, $H_i^{\SCL}$ tends to be close to $H_i^{\ML}$. From a practical point of view, the simulation results in \cite{Tal15} tell us that when the list size is $L=32$, the performance of the SCL decoder is already very close to the ML decoder for the Binary-Input AWGN channel with $E_b/N_0>1.5\dB$. Therefore, for list size $L=32$, $H_i^{\SCL}$ is very close to $H_i^{\ML}$ for a wide range of channels. This further indicates that for some values of $i$, the gap between $H_i^{\SCL}$ and $H_i$ might not be negligible.
	
	Next let us consider how to improve the polar code construction to compensate for the gap between $H_i^{\SCL}$ and $H_i$. Observe that $U_i$ with smaller index $i$ have more future frozen bits in front of it, so intuitively, the gap $H_i-H_i^{\SCL}$ tends to be larger for smaller value of $i$. This suggests that the reliability of $U_i$ with relatively small index are more likely to be underestimated by the standard polar code construction. As a consequence, in the optimal polar code construction for the SCL decoder, more information bits should be assigned to the minus branch\footnote{Recall the definition of minus branch at the end of Section~\ref{sect:bg}: By minus branch we simply mean $(U_0,U_1,\dots,U_{n/2-1})$, the first half of $(U_0,U_1,\dots,U_{n-1})$.} than in the standard polar code construction. In the next section, we will develop a dynamic programming algorithm to find the optimal number of information bits in the minus branch for the SCL decoder.
	
	\section{A dynamic programming algorithm to construct polar codes for SCL decoder} \label{sect:DP}
	
	Similarly to RM codes and the standard polar codes, the generator matrices of our DP-polar codes are also submatrices of $G_m:=\begin{bmatrix}
		1 & 0 \\
		1 & 1
	\end{bmatrix}^{\otimes m}$. Among all the $(n=2^m,k)$ codes whose generator matrices consist of $k$ row vectors of $G_m$, we want to find the one with the best performance under the SCL decoder. A naive way to do so is to test the decoding error probability of all the ${n \choose k}$ candidates under the SCL decoder. However, the complexity of this naive method is too high for large values of $n$. Therefore, we propose the following dynamic programming method to find the best $k$ rows of $G_m$ under the SCL decoder.
	
	Although our DP-polar code construction works for any class of communication channels, below we will focus on the construction for binary-input AWGN channel because this is the most important communication channel in practice. 
	
	\subsection{Construct polar codes from the minus array}
	
	Our code construction depends on three parameters---the code length $n$, the code dimension $k$, and the list size $L$. Following the binary tree decomposition in Fig.~\ref{fig:fceng}, we use $\minus_L(n,k)$ to denote the number of information bits assigned to the minus branch when we construct an $(n,k)$ polar code that is optimized for list size $L$.
	For a fixed value of $L$, if we know $\minus_L(n,k)$ for all possible pairs of $(n,k)$, then we can construct polar codes with any code length and code dimension according to the binary tree decomposition. More precisely, constructing an $(n,k)$ polar code amounts to finding $\cA$, the index set of the $k$ information bits. Let $k^-=\minus_L(n,k)$ and $k^+=k-k^-$. Then we know that $k^-$ information bits are assigned to the minus branch, i.e., there are $k^-$ information bits in $(U_0,U_1,\dots,U_{n/2-1})$. Similarly, there are $k^+$ information bits in $(U_{n/2},U_{n/2+1},\dots,U_{n-1})$. Next we look at the values of $k^{--}=\minus_L(n/2,k^-), k^{-+}=k^- - k^{--}, k^{+-}=\minus_L(n/2,k^+), k^{++}=k^+ - k^{+-}$. According to the binary tree decomposition, $k^{--}$ is the number of information bits in $(U_0,\dots,U_{n/4-1})$; $k^{-+}$ is the number of information bits in $(U_{n/4},\dots,U_{n/2-1})$; $k^{+-}$ is the number of information bits in $(U_{n/2},\dots,U_{3n/4-1})$; $k^{++}$ is the number of information bits in $(U_{3n/4},\dots,U_{n-1})$. This decomposition continues until we obtain $\{k^s:s\in\{+,-\}^m\}$, where the value of each $k^s$ indicates whether the corresponding $U_i$ is an information bit or a frozen bit. More precisely, for every $0\le i\le 2^m-1$, let $s(i)\in\{+,-\}^m$ be the sequence obtained by replacing ``$0$" with ``$-$" and replacing ``$1$" with ``$+$" in the $m$-digit binary expansion of $i$. Then $U_i$ is an information bit if $k^{s(i)}=1$, and $U_i$ is a frozen bit if $k^{s(i)}=0$. Note that $k^s$ is either $0$ or $1$ for all $s\in\{+,-\}^m$. In this way, we obtain an $(n,k)$ polar code that is optimized for the SCL decoder with list size $L$. This construction method is summarized in Algorithm~\ref{algo:1} below.

	\begin{algorithm}[H]
		\DontPrintSemicolon
		\caption{\texttt{MinusConstruct}$(n,k,\{\minus_L(n',k'): n'\le n,0\le k'\le n'\})$}
		\label{algo:1}
		\KwIn{code length $n\ge 2$, code dimension $k$, and the minus array $\{\minus_L(n',k'): n'\le n,0\le k'\le n'\}$}
		\KwOut{the index set $\cA$ of the information bits}
		
		$m\gets \log_2(n)$
		
		$\cA\gets \emptyset$
		\Comment{Initialize $\cA$ as the empty set.}
		
		\For{$j=0,1,\dots,m-1$}
		{ \For{$s\in\{+,-\}^j$}
			{
				\Comment{$s$ is the empty string when $j=0$. In this case, $k^s=k$.}
				
				$k^{s-}\gets \minus_L(n/2^j,k^s)$
				
				\Comment{$s-$ is obtained from appending a ``$-$" at the end of the string $s$}
				
				$k^{s+}\gets k^s - k^{s-}$
				
				\Comment{$s+$ is obtained from appending a ``$+$" at the end of the string $s$}
			}
		}
		
		\For{$i=0,1,2,\dots,n-1$}
		{
			\If{$k^{s(i)}=1$}
			{
				$\cA\gets \cA\cup\{i\}$
			}
		}
		
		\Return $\cA$ 
		
	\end{algorithm}

	\begin{table}[H]
		\centering
		\large
		\begin{tabular}{c|cccccccccccccccccc}
			\hline
			\diagbox{$n$}{$k$} &0 &1 &2 &3 &4 &5 &6 &7 &8 &9 &10 &11 &12 &13 &14 &15 &16\\
			\hline
			
			2 &0 &0 &1\\
			
			4 &0 &0 &0 &1 &2\\
			
			8 &0 &0 &0 &0 &1 &1 &2 &3 &4\\
			
			16 &0 &0 &0 &0 &1 &1 &1 &2 &2 &2 &3 &4 &4 &6 &7 &7 &8\\
			\hline
		\end{tabular}
		\caption{An example of the minus array with $n\le 16$ and $0\le k\le n$.}
		\label{table:example}
	\end{table}
	
	As a concrete example, we show how to construct an $(n=16,k=11)$ code from the minus array in Table~\ref{table:example}. The first step is to calculate $k^-$ and $k^+$:
	$$
	k^- = \minus_{L}(16,11) = 4, \quad\quad 
	k^+= 11- k^- = 7.
	$$
	The next step is to calculate $k^{--},k^{-+},k^{+-}, k^{++}$:
	\begin{align*}
		& k^{--}=\minus_L(8,k^-)= 1, \quad 
		k^{-+}=k^- - k^{--} = 3, \\
		& k^{+-}= \minus_L(8,k^+) = 3, \quad
		k^{++} = k^+ - k^{+-} = 4 .
	\end{align*}
	Next we calculate $k^s, s\in\{+,-\}^3$:
	\begin{align*}
		k^{---}=\minus_L(4,k^{--})=0, \quad
		k^{--+}=k^{--} - k^{---} = 1, \\
		k^{-+-}=\minus_L(4,k^{-+})= 1, \quad
		k^{-++}=k^{-+}- k^{-+-}=2, \\
		k^{+--}=\minus_L(4,k^{+-})=1, \quad
		k^{+-+}=k^{+-}-k^{+--}=2, \\
		k^{++-}=\minus_L(4,k^{++})=2, \quad
		k^{+++}=k^{++}- k^{++-}=2 .
	\end{align*}
	Finally, we obtain
	\begin{align*}
		k^{----}=0, \quad k^{---+}=0, \quad k^{--+-}=0, \quad k^{--++}=1, \\
		k^{-+--}=0, \quad k^{-+-+}=1, \quad k^{-++-}=1, \quad k^{-+++}=1, \\
		k^{+---}=0, \quad k^{+--+}=1, \quad k^{+-+-}=1, \quad k^{+-++}=1, \\
		k^{++--}=1, \quad k^{++-+}=1, \quad k^{+++-}=1, \quad k^{++++}=1,
	\end{align*}
	so the index set of the information bits is $\cA = \{3,5,6,7,9,10,11,12,13,14,15\}$. This completes the construction of the $(n=16,k=11)$ code. Note that this $(16,11)$ code is the same as the code in Fig.~\ref{fig:fceng}.

	\subsection{Obtain the minus array by dynamic programming}
	
	Let's say we want to construct polar codes with code length up to $N$, where $N$ is a power of $2$.
	In our implementation, we take $N=1024$. For a given list size $L$, we calculate the minus array $\minus_L(n,k)$ from small values of $n$ to large values of $n$. For each fixed value of $n$, we calculate $\minus_L(n, k)$ from $k = 0$ to $k = n$. As a consequence, when we calculate the value of $\minus_L(n,k)$, we already know $\{\minus_L(n',k'): n'\le n/2,0\le k'\le n'\}$. It is easy to see that the range of $\minus_L(n,k)$ is 
	\begin{equation} \label{eq:range}
		\max(0, k - n/2) \le \minus_L(n,k) \le \min(k,n/2).
	\end{equation}
	For each choice of $\minus_L(n,k)=k^-$, we construct the corresponding polar codes as follows: Let $\cA_1$ be the index set of information bits returned by Algorithm~\ref{algo:1} when the inputs to the algorithm are code length $n/2$, code dimension $k^-$, and the minus array $\{\minus_L(n',k'): n'\le n/2,0\le k'\le n'\}$. Let $\cA_2$ be the index set of information bits returned by Algorithm~\ref{algo:1} when the inputs to the algorithm are code length $n/2$, code dimension $k-k^-$, and the minus array $\{\minus_L(n',k'): n'\le n/2,0\le k'\le n'\}$. Note that both $\cA_1$ and $\cA_2$ are subsets of $\{0,1,2,\dots,n/2-1\}$. Next define $\cA_2+n/2:=\{i+n/2:i\in\cA_2\}$ and define $\cA=\cA_1 \cup (\cA_2+n/2)$. The set $\cA$ is the index set of information bits in the code construction corresponding to the choice $\minus_L(n,k)=k^-$. For each $k^-$ between $\max(0, k - n/2)$ and $\min(k,n/2)$, we test the performance of its corresponding code construction under the SCL decoder with list size $L$. The testing channel is the binary-input AWGN channel with $E_b/N_0=2\dB$. We set the value of $\minus_L(n,k)$ to be the optimal choice of $k^-$ which minimizes the decoding error probability under the SCL decoder with list size $L$. 
	This dynamic programming approach is summarized in Algorithm~\ref{algo:2} below.
	The function $\texttt{DecodingError}(n,k, \cA, L)$ in line 11 of Algorithm~\ref{algo:2} estimates the decoding error probability of the $(n,k)$ polar code under the SCL decoder with list size $L$, where $\cA$ is the index set of information bits. In our implementation, we perform $10^5$ rounds of Monte Carlo simulations over the binary-input AWGN channel with $E_b/N_0=2\dB$ to estimate the decoding error probability of the SCL decoder.

	\begin{algorithm}[H]
		\DontPrintSemicolon
		\caption{\texttt{CalculateMinusArray}}
		\label{algo:2}
		\KwIn{upper bound $N$ on the code length ($N$ is a power of $2$), and the list size $L$}
		\KwOut{the minus array $\{\minus_L(n,k):n\le N,0\le k\le n\}$}
		\For{$n = 2, 4, 8, \dots, N$}{
			\For{$k = 0, 1, \dots, n$}{
				$s \gets \max(0, k-n/2)$
				
				$t \gets \min(k,n/2)$

				$err_{\min} \gets 1$
				\Comment{Initialize the minimum decoding error probability as $1$}
				
				\For{$k^- = s, s+1,s+2,\dots, t$}{
					
					$\cA_1 \gets \texttt{MinusConstruct}(n/2, k^-, \{\minus_L(n',k'): n'\le n/2,0\le k'\le n'\})$
					
					\Comment{The function \texttt{MinusConstruct} is defined in Algorithm~\ref{algo:1}}
					
					$\cA_2 \gets \texttt{MinusConstruct}(n/2, k- k^-, \{\minus_L(n',k'): n'\le n/2,0\le k'\le n'\})$
					
					$\cA \gets \cA_1 \cup (\cA_2+n/2)$
					
					$err \gets \texttt{DecodingError}(n,k, \cA, L)$
					
					\Comment{See the main text about how the function $\texttt{DecodingError}(n,k, \cA, L)$ works.}
					
					\If{$err< err_{\min}$}{
						$err_{\min} \gets err$
						
						$\minus_L(n,k) \gets k^-$ 
					}						
				}

			}
		}

		\Return $\{\minus_L(n,k):n\le N,0\le k\le n\}$
		
	\end{algorithm}
	
	\begin{remark}
		The range in \eqref{eq:range} is too large for practical implementation. We can use the continuity of $\minus_L(n,k)$ to shrink the range of search. Intuitively, we would expect that $\minus_L(n,k-1)$ is close to $\minus_L(n,k)$. Since we calculate the minus array from small values of $k$ to large values of $k$, we already know the value of $\minus_L(n,k-1)$ when we calculate $\minus_L(n,k)$. Therefore, we only need to search in a small neighborhood of $\minus_L(n,k-1)$ in order to determine the value of $\minus_L(n,k)$. In our implementation, we only test $7$ values in the set $\{\minus_L(n,k-1) -2, \minus_L(n,k-1) -1, \dots, \minus_L(n,k-1)+3, \minus_L(n,k-1)+4\}$ for the calculation of each $\minus_L(n,k)$.
	\end{remark}
	
	\subsection{Connection with RM codes}
	The generator matrices of standard polar codes, DP-polar codes and RM codes are all submatrices of the same square matrix $G_m:=\begin{bmatrix}
		1 & 0 \\
		1 & 1
	\end{bmatrix}^{\otimes m}$. These three classes of codes pick different sets of row vectors from $G_m$ to form their generator matrices. More precisely, standard polar codes pick the index set of information bits according to the conditional expectation $H_i$ defined in \eqref{eq:hi}; DP-polar codes pick the index set of information bits according to the minus array; and RM codes pick the index set according to the Hamming weight of the binary expansion of the row index, as described in Section~\ref{sect:RM}. This allows us to measure the similarity between these three classes of codes as follows: Let $n=2^m$ be the code length and let 
	$$
	k=\binom{m}{0}+\binom{m}{1}+\binom{m}{2}+\dots+\binom{m}{r}
	$$ 
	be the code dimension\footnote{We restrict ourselves to these specific choices of code dimension because the dimension of RM codes can only take this form.}, where $r$ is some positive integer that is smaller than $m$. For the $(n,k)$ standard polar code, we denote its index set of information bits as $\cA_{\polar}$. For the $(n,k)$ DP-polar code, we denote its index set of information bits as $\cA_{\DP}$. For the $(n,k)$ RM code, we denote its index set of information bits as $\cA_{\RM}$. We define the similarity between the $(n,k)$ DP-polar code and the $(n,k)$ standard polar code as
	$$
	S_{\polar}:=\frac{|\cA_{\DP}\cap\cA_{\polar}|}{k} ,
	$$
	where $|\cA|$ denotes the size of the set $\cA$. The similarity between the $(n,k)$ DP-polar code and the $(n,k)$ RM code is defined as
	$$
	S_{\RM}:=\frac{|\cA_{\DP}\cap\cA_{\RM}|}{k} .
	$$
	By definition, $S_{\polar}=1$ implies that the $(n,k)$ DP-polar code is the same as the standard polar code. Similarly, $S_{\RM}=1$ implies that the $(n,k)$ DP-polar code is the same as the RM code.
	
	Since our DP-polar code construction depends on the list size $L$, the similarity measures $S_{\polar}$ and $S_{\RM}$ are both functions of $L$. In Tables~\ref{table:construction}--\ref{table:construction3} below, we list how the values of $S_{\polar}$ and $S_{\RM}$ change with the list size $L$. As we can see from the tables, $S_{\polar}$ always decreases with $L$, and $S_{\RM}$ always increases with $L$. When $L=1$, the DP-polar code is (almost) the same as the standard polar code. When $L$ is large enough, e.g., when $L=128$, the DP-polar code is the same as the RM code. This tells us that when we increase $L$ from $1$ to infinity, the DP-polar code gradually changes from the standard polar code to the RM code.
	
	The construction of DP-polar codes aims to optimize its performance under the SCL decoder with a given list size $L$. When $L$ goes to infinity, the SCL decoder is the same as the ML decoder. Therefore, the fact that DP-polar codes converge to RM codes when $L\to\infty$ suggests that RM codes have the best performance under the ML decoder among all the codes whose generator matrices are formed of row vectors of $G_m=\begin{bmatrix}
		1 & 0 \\
		1 & 1
	\end{bmatrix}^{\otimes m}$. In particular, this suggests that RM codes have better performance than polar codes with the same parameters under the ML decoder. Note that this observation has already been made in \cite{Mondelli14,Ye20}. The discussion above provides one more indirect evidence to this claim.

	\begin{table}[H]
		\centering
		\begin{tabular}{c|cccccccc}
			\hline
			$L$ &1 &2 &4 &8 &16 &32 &64 &128  \\
			\hline
			$S_{\polar}$  &1.00  &0.93  &0.93  &0.86  &0.83  &0.83  &0.83  &0.83  \\
			\hline
			$S_{\RM}$     &0.83  &0.90  &0.90  &0.97  &1.00  &1.00  &1.00  &1.00  \\ 
			\hline
			
		\end{tabular}
		\caption{The values of $S_{\polar}$ and $S_{\RM}$ change with the list size $L$. Here we take $n = 128$ and $k = 29$, corresponding to the second-order RM code.}
		\label{table:construction}
	\end{table}
	
	\begin{table}[H]
		\centering
		\begin{tabular}{c|cccccccc}
			\hline
			$L$ &1 &2 &4 &8 &16 &32 &64 &128  \\
			\hline
			$S_{\polar}$    &0.97  &0.97  &0.95  &0.94  &0.92  &0.92  &0.91  &0.91\\
			\hline
			$S_{\RM}$       &0.91  &0.94  &0.94  &0.95  &0.98  &0.98  &0.98  &1.00\\ 
			\hline
			
		\end{tabular}
		\caption{The values of $S_{\polar}$ and $S_{\RM}$ change with the list size $L$. Here we take $n = 128$ and $k = 64$, corresponding to the third-order RM code.}
		\label{table:construction2}
	\end{table}
	
	\begin{table}[H]
		\centering
		\begin{tabular}{c|cccccccc}
			\hline
			$L$ &1 &2 &4 &8 &16 &32 &64 &128  \\
			\hline
			$S_{\polar}$    &1.00  &0.99  &0.97  &0.96  &0.95  &0.95  &0.95  &0.95\\
			\hline
			$S_{\RM}$       &0.95  &0.96  &0.98  &0.99  &0.99  &0.99  &0.99  &1.00 \\ 
			\hline
			
		\end{tabular}
		\caption{The values of $S_{\polar}$ and $S_{\RM}$ change with the list size $L$. Here we take $n = 128$ and $k = 99$, corresponding to the fourth-order RM code.}
		\label{table:construction3}
	\end{table}

	\section{Simulation results} \label{sect:simu}

	\begin{figure}
		\centering
		\begin{subfigure}{0.42\linewidth} 
			\centering
			\includegraphics[width=\linewidth]{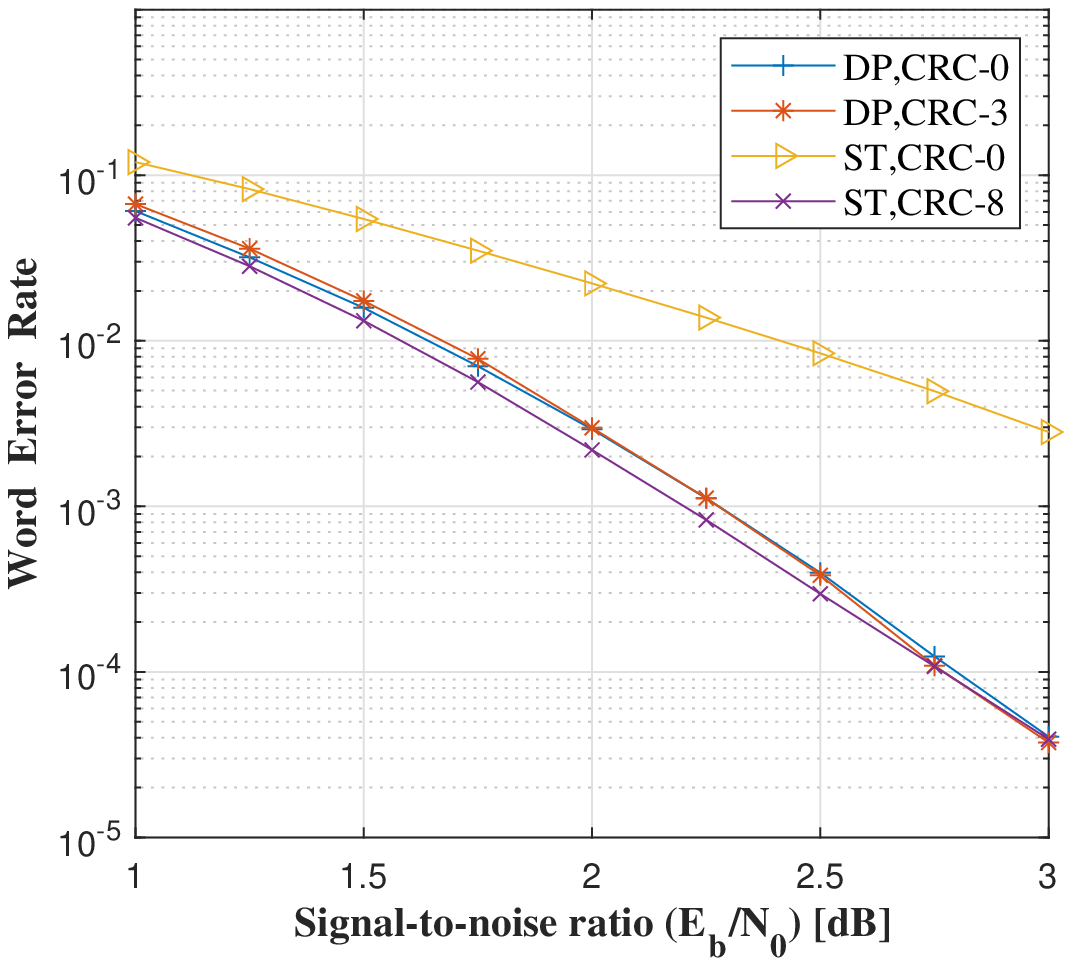}
			\caption{length 256, dimension 77}
		\end{subfigure}
		~\hspace*{0.2in}
		\begin{subfigure}{0.42\linewidth}
			\centering
			\includegraphics[width=\linewidth]{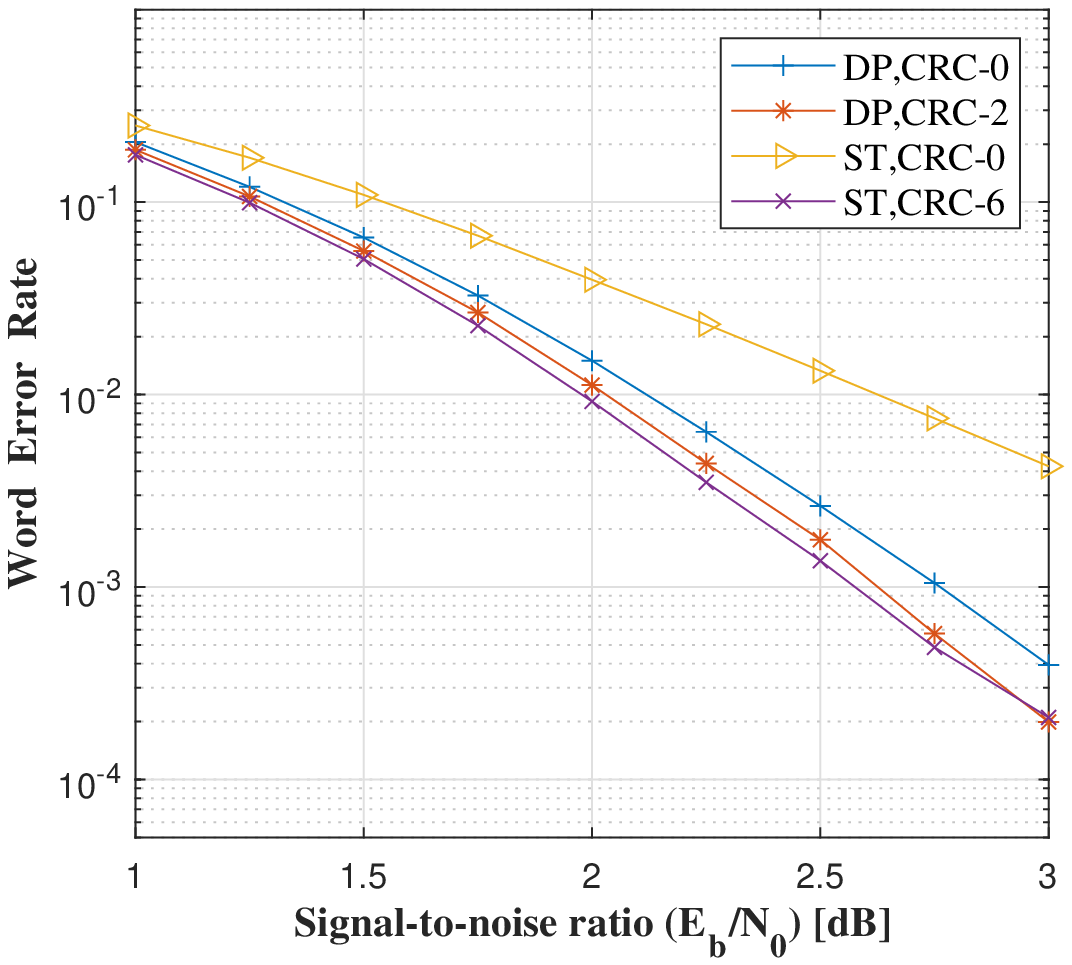}
			\caption{length 256, dimension 128}
		\end{subfigure}

		\vspace*{0.1in}
		
		\begin{subfigure}{0.42\linewidth} 
			\centering
			\includegraphics[width=\linewidth]{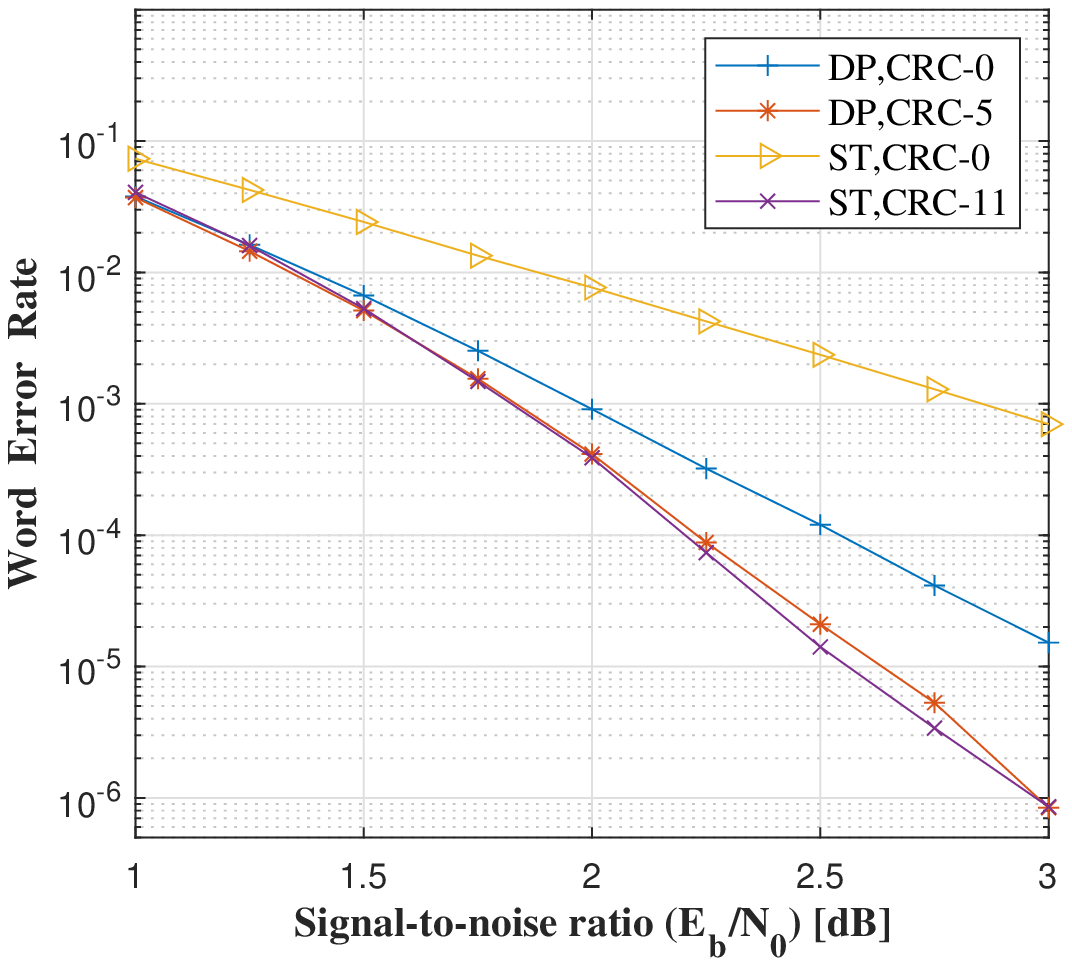}
			\caption{length 512, dimension 154}
		\end{subfigure}
		~\hspace*{0.2in}
		\begin{subfigure}{0.42\linewidth}
			\centering
			\includegraphics[width=\linewidth]{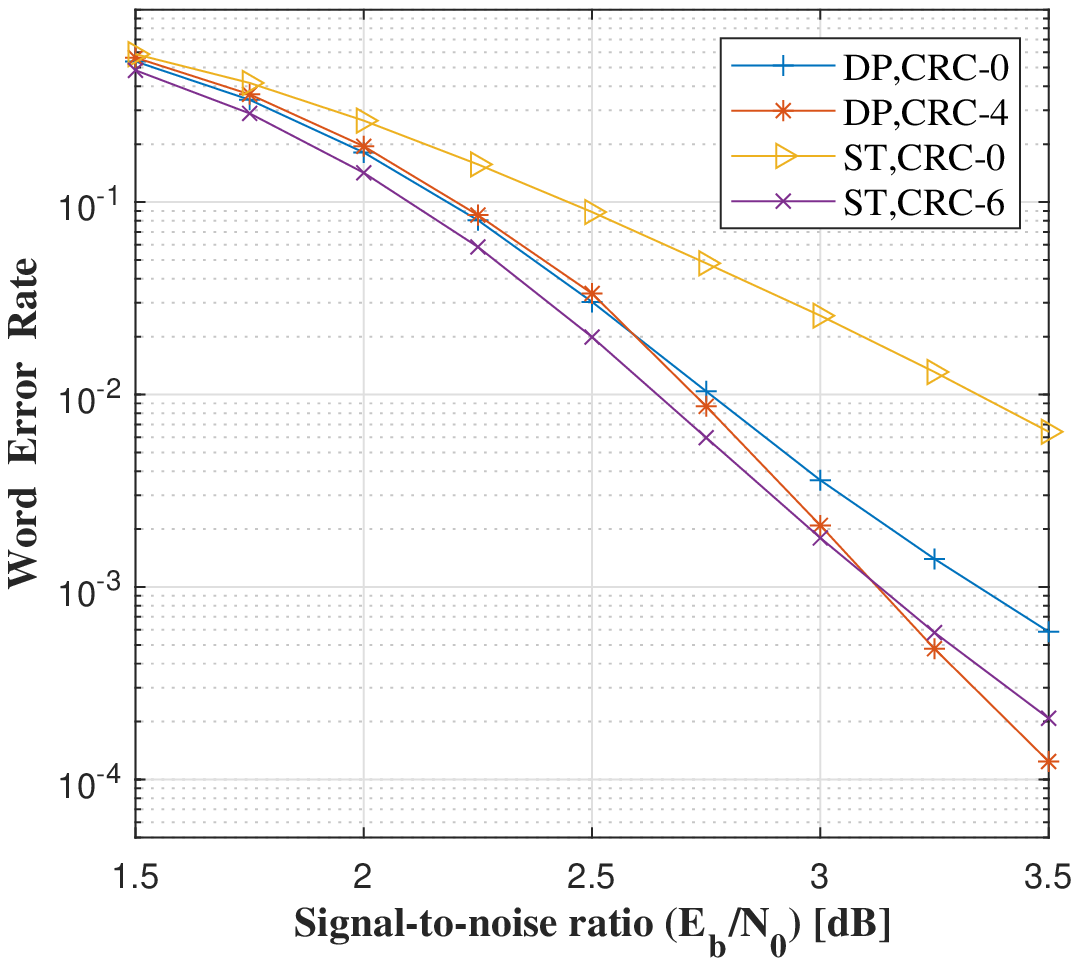}
			\caption{length 512, dimension 358}
		\end{subfigure}
		
		\vspace*{0.1in}
		\begin{subfigure}{0.42\linewidth} 
			\centering
			\includegraphics[width=\linewidth]{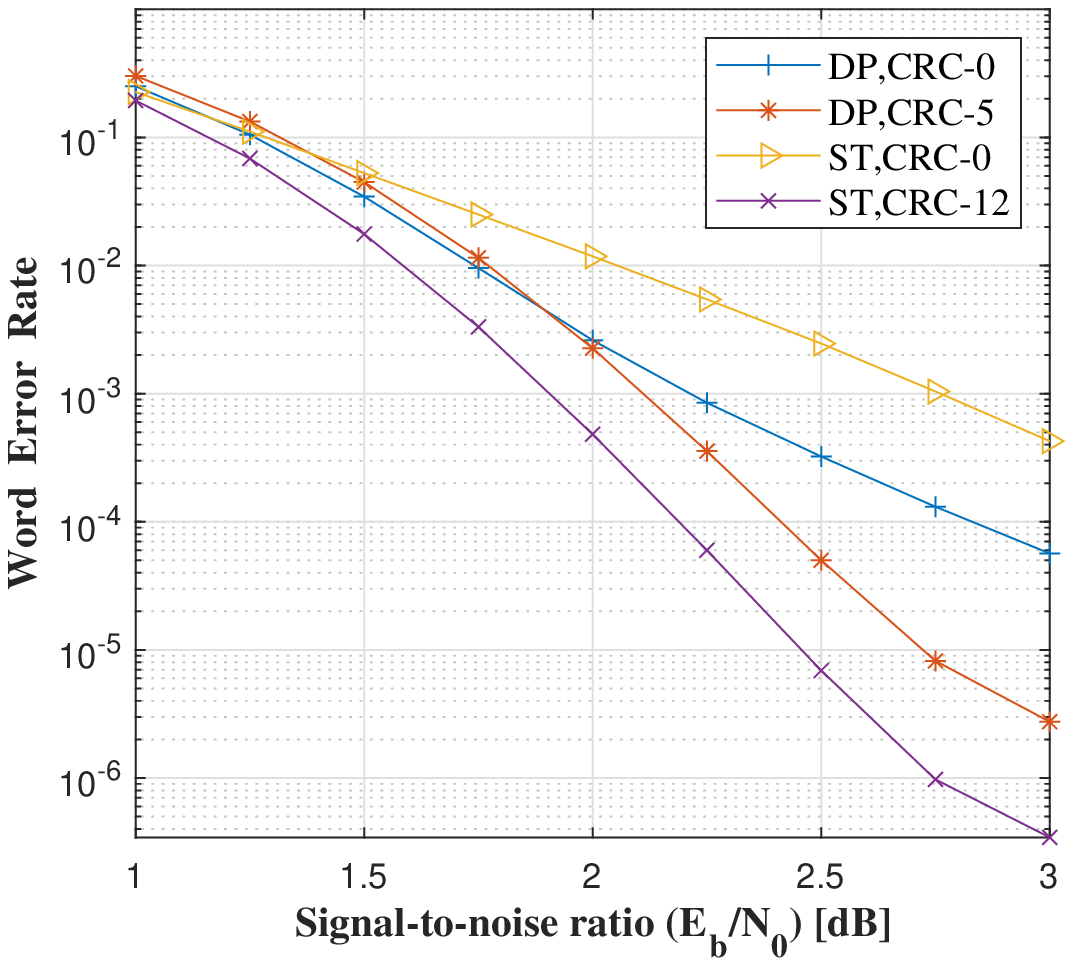}
			\caption{length 1024, dimension 512}   
		\end{subfigure}
		~\hspace*{0.2in}
		\begin{subfigure}{0.42\linewidth} 
			\centering
			\includegraphics[width=\linewidth]{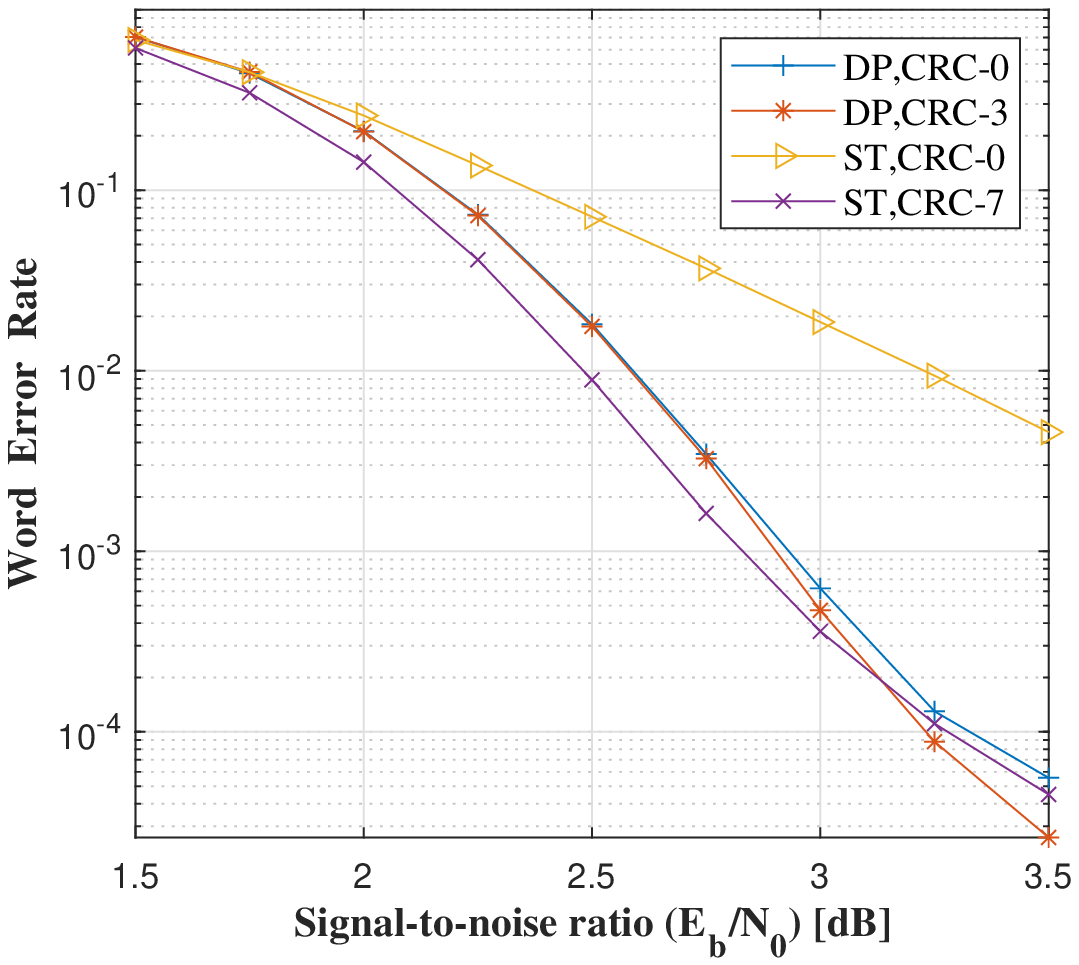}
			\caption{length 1024, dimension 717}   
		\end{subfigure}
		
		\caption{Comparison between DP-polar codes and standard polar codes over the binary-input AWGN channel. We use SCL decoder with list size $32$ in all the plots. The legend ``DP" refers to DP-polar codes, and ``ST" refers to standard polar codes. ``CRC-0" means that we do not use CRC. ``CRC-$\ell_c$" with a nonzero $\ell_c$ means that $\ell_c$ is the optimal CRC length that minimizes the decoding error probability.}
		\label{fig:cpAWGN}
	\end{figure}

	In this section, we provide simulation results to compare the performance of the DP-polar codes with the standard polar codes under the SCL decoder with list size $32$. For this choice of list size, the minus array of our DP-polar code construction is given in Appendix~\ref{ap:ma}. All the DP-polar codes in Fig.~\ref{fig:cpAWGN} are constructed from this minus array.
	
	Similarly to the standard polar codes, we can also add CRC to the DP-polar code construction and the corresponding SCL decoder. In order to add a length $\ell_c$ CRC to an $(n,k)$ DP-polar code, we first construct an $(n,k+\ell_c)$ DP-polar code from the minus array. Then we divide these $k+\ell_c$ information bits into two parts: The first $k$ bits carry the message, and the last $\ell_c$ bits are (random) parity checks of the first $k$ bits. In the SCL decoding procedure, the CRC allows us to eliminate incorrect candidates in the list.

	
	The simulation results in Fig.~\ref{fig:cpAWGN} include various choices of code length $n\in\{256,512,1024\}$ and code rates $k/n \in\{0.3,0.5,0.7\}$. In each plot, there are four curves, where the legend ``DP" refers to DP-polar codes, and the legend ``ST" refers to standard polar codes. Moreover, ``CRC-0" in the legend means that we do not use CRC. ``CRC-$\ell_c$" with a nonzero $\ell_c$ means that $\ell_c$ is the optimal CRC length that minimizes the decoding error probability at $E_b/N_0=2.5\dB$.
	
	From these simulation results, we can see that when there is no CRC, our DP-polar codes consistently outperform the standard polar codes by $0.3$--$1$dB. When we use the optimal length of CRC on both codes, the performance of DP-polar codes is similar to the standard polar codes for most choices of parameters. 
	As a final remark, since we use exactly the same decoder for both the DP-polar codes and the standard polar codes, the decoding time of these two classes of codes are more or less the same.

	\appendices
	
	\section{The minus array for $L=32$} \label{ap:ma}

	\begin{table}[H]
		\centering
		
		\resizebox{\columnwidth}{!}{
			\begin{tabular}{c|c|cccccccccccccccc}
				\hline
				$n$ &\diagbox{$k$}{$\Delta$} &0 &1 &2 &3 &4 &5 &6 &7 &8 &9 &10 &11 &12 &13 &14 &15\\
				\hline
				\multirow{ 1}*{   2} &1     &0     &1     \\ \cline{2-18}

				\hline
				\multirow{ 1}*{   4} &1     &0     &0     &1     &2     \\ \cline{2-18}

				\hline
				\multirow{ 1}*{   8} &1     &0     &0     &0     &1     &1     &2     &3     &4     \\ \cline{2-18}

				\hline
				\multirow{ 1}*{  16} &1     &0     &0     &0     &1     &1     &1     &2     &1     &2     &4     &4     &4     &5     &6     &7     &8     \\ \cline{2-18}

				\hline
				\multirow{ 2}*{  32} &1     &0     &0     &0     &1     &0     &1     &1     &3     &1     &3     &1     &4     &3     &4     &5     &5     \\ \cline{2-18}
				~                    &17    &5     &6     &7     &7     &6     &10    &10    &9     &10    &11    &11    &12    &13    &14    &15    &16    \\ \cline{2-18}

				\hline
				\multirow{ 4}*{  64} &1     &0     &0     &0     &0     &0     &0     &1     &2     &1     &2     &3     &3     &1     &2     &4     &3     \\ \cline{2-18}
				~                    &17    &5     &5     &5     &5     &5     &6     &7     &7     &8     &8     &8     &8     &8     &11    &9     &9     \\ \cline{2-18}
				~                    &33    &11    &11    &13    &13    &12    &14    &15    &16    &16    &16    &16    &18    &18    &19    &19    &19    \\ \cline{2-18}
				~                    &49    &22    &22    &23    &23    &23    &23    &25    &25    &26    &26    &28    &29    &30    &31    &31    &32    \\ \cline{2-18}

				\hline
				\multirow{ 8}*{ 128} &1     &0     &0     &0     &0     &1     &0     &1     &1     &1     &3     &3     &1     &2     &1     &2     &1     \\ \cline{2-18}
				~                    &17    &2     &3     &4     &3     &4     &4     &3     &4     &4     &6     &6     &6     &7     &7     &7     &8     \\ \cline{2-18}
				~                    &33    &9     &7     &8     &7     &8     &7     &9     &8     &9     &9     &10    &10    &12    &12    &13    &13    \\ \cline{2-18}
				~                    &49    &13    &14    &13    &15    &14    &13    &15    &15    &16    &16    &18    &18    &19    &19    &21    &21    \\ \cline{2-18}
				~                    &65    &21    &21    &21    &22    &22    &23    &23    &23    &23    &24    &25    &25    &26    &27    &28    &29    \\ \cline{2-18}
				~                    &81    &29    &29    &29    &29    &31    &31    &32    &32    &32    &33    &34    &37    &37    &37    &38    &39    \\ \cline{2-18}
				~                    &97    &40    &41    &42    &43    &43    &43    &43    &46    &46    &47    &48    &47    &47    &48    &50    &50    \\ \cline{2-18}
				~                    &113   &50    &51    &53    &54    &54    &56    &56    &57    &58    &59    &60    &60    &61    &63    &63    &64    \\ \cline{2-18}

				\hline
				\multirow{16}*{ 256} &1     &0     &0     &0     &0     &0     &0     &0     &0     &1     &2     &1     &1     &1     &1     &1     &2     \\ \cline{2-18}
				~                    &17    &1     &2     &3     &4     &5     &5     &3     &1     &1     &2     &4     &4     &4     &5     &4     &4     \\ \cline{2-18}
				~                    &33    &5     &4     &5     &5     &6     &7     &5     &6     &7     &5     &7     &8     &7     &8     &7     &8     \\ \cline{2-18}
				~                    &49    &7     &8     &9     &7     &8     &8     &9     &9     &9     &10    &10    &9     &10    &11    &10    &10    \\ \cline{2-18}
				~                    &65    &11    &10    &10    &11    &11    &12    &14    &14    &15    &14    &15    &15    &14    &14    &15    &16    \\ \cline{2-18}
				~                    &81    &16    &17    &17    &17    &17    &18    &18    &18    &19    &22    &23    &23    &23    &22    &23    &23    \\ \cline{2-18}
				~                    &97    &25    &25    &25    &25    &25    &25    &25    &25    &25    &27    &27    &27    &27    &27    &27    &28    \\ \cline{2-18}
				~                    &113   &28    &31    &31    &29    &31    &31    &31    &31    &31    &31    &32    &36    &36    &36    &36    &36    \\ \cline{2-18}
				~                    &129   &36    &36    &38    &38    &40    &40    &40    &40    &40    &41    &41    &41    &42    &44    &44    &44    \\ \cline{2-18}
				~                    &145   &46    &46    &47    &46    &47    &49    &49    &51    &51    &51    &54    &54    &54    &54    &54    &56    \\ \cline{2-18}
				~                    &161   &56    &58    &58    &58    &58    &58    &58    &60    &60    &60    &62    &62    &62    &64    &66    &66    \\ \cline{2-18}
				~                    &177   &66    &66    &66    &67    &69    &68    &69    &69    &69    &71    &71    &71    &71    &73    &73    &74    \\ \cline{2-18}
				~                    &193   &75    &76    &77    &76    &77    &78    &81    &84    &83    &84    &84    &84    &84    &88    &89    &88    \\ \cline{2-18}
				~                    &209   &89    &89    &89    &91    &91    &94    &94    &94    &94    &95    &95    &96    &99    &99    &102   &101   \\ \cline{2-18}
				~                    &225   &102   &103   &102   &103   &103   &103   &105   &109   &109   &108   &109   &110   &110   &112   &113   &114   \\ \cline{2-18}
				~                    &241   &117   &115   &116   &118   &119   &120   &120   &120   &122   &122   &124   &124   &125   &126   &127   &128   \\ \cline{2-18}
				\hline
			\end{tabular}
		}		
		\caption{Minus array for $L = 32, n = 2,4,\dots, 256$ and $1\le k\le n$. The value in the table is $\minus_{32}(n,k+\Delta)$.}
		\label{table:minus_n2-256}
	\end{table}
	
	\begin{table}[H]
		\centering
		\resizebox{\columnwidth}{!}{
			\begin{tabular}{c|c|cccccccccccccccc}
				\hline
				$n$ &\diagbox{$k$}{$\Delta$} &0 &1 &2 &3 &4 &5 &6 &7 &8 &9 &10 &11 &12 &13 &14 &15\\
				
				\hline
				\multirow{32}*{ 512} &1     &0     &0     &0     &0     &1     &0     &0     &1     &1     &1     &2     &3     &1     &1     &1     &1     \\ \cline{2-18}
				~                    &17    &2     &1     &1     &3     &4     &5     &5     &5     &3     &2     &1     &4     &5     &4     &2     &3     \\ \cline{2-18}
				~                    &33    &2     &2     &3     &2     &3     &2     &3     &4     &4     &3     &4     &4     &2     &3     &4     &3     \\ \cline{2-18}
				~                    &49    &4     &5     &4     &5     &4     &5     &5     &6     &5     &6     &5     &6     &7     &5     &6     &4     \\ \cline{2-18}
				~                    &65    &5     &6     &7     &8     &6     &7     &7     &6     &7     &8     &7     &7     &6     &7     &8     &8     \\ \cline{2-18}
				~                    &81    &9     &10    &9     &10    &10    &9     &10    &11    &10    &11    &11    &11    &11    &11    &11    &11    \\ \cline{2-18}
				~                    &97    &11    &12    &15    &15    &15    &15    &15    &13    &13    &13    &15    &15    &16    &15    &15    &16    \\ \cline{2-18}
				~                    &113   &17    &17    &17    &17    &17    &18    &18    &17    &18    &18    &18    &19    &18    &19    &19    &19    \\ \cline{2-18}
				~                    &129   &20    &20    &24    &24    &24    &24    &24    &24    &24    &24    &24    &24    &24    &24    &24    &26    \\ \cline{2-18}
				~                    &145   &26    &25    &24    &24    &24    &25    &26    &25    &27    &26    &27    &29    &29    &29    &29    &29    \\ \cline{2-18}
				~                    &161   &29    &29    &29    &29    &30    &30    &30    &30    &30    &31    &32    &32    &31    &32    &32    &34    \\ \cline{2-18}
				~                    &177   &34    &36    &36    &35    &36    &36    &36    &39    &39    &40    &42    &42    &42    &45    &45    &45    \\ \cline{2-18}
				~                    &193   &43    &42    &45    &45    &45    &45    &47    &45    &45    &47    &47    &46    &47    &46    &47    &48    \\ \cline{2-18}
				~                    &209   &47    &48    &49    &49    &49    &52    &52    &54    &54    &54    &54    &54    &53    &54    &54    &55    \\ \cline{2-18}
				~                    &225   &54    &55    &56    &60    &60    &60    &60    &61    &64    &64    &64    &64    &64    &64    &64    &65    \\ \cline{2-18}
				~                    &241   &64    &65    &66    &64    &64    &65    &69    &69    &69    &69    &69    &69    &69    &70    &69    &72    \\ \cline{2-18}
				~                    &257   &72    &72    &72    &73    &71    &72    &73    &74    &73    &72    &73    &74    &78    &78    &79    &78    \\ \cline{2-18}
				~                    &273   &78    &79    &81    &81    &83    &84    &84    &85    &84    &83    &83    &83    &84    &85    &87    &87    \\ \cline{2-18}
				~                    &289   &87    &88    &87    &88    &86    &87    &88    &87    &88    &89    &92    &96    &94    &96    &96    &96    \\ \cline{2-18}
				~                    &305   &96    &99    &99    &99    &102   &101   &102   &103   &104   &105   &105   &104   &103   &104   &105   &105   \\ \cline{2-18}
				~                    &321   &105   &108   &109   &110   &111   &111   &111   &112   &111   &112   &110   &111   &112   &113   &112   &113   \\ \cline{2-18}
				~                    &337   &116   &120   &118   &119   &120   &119   &120   &122   &122   &122   &122   &125   &127   &127   &127   &127   \\ \cline{2-18}
				~                    &353   &130   &130   &130   &130   &132   &132   &134   &134   &134   &134   &135   &135   &137   &140   &140   &140   \\ \cline{2-18}
				~                    &369   &140   &139   &140   &140   &140   &140   &144   &144   &143   &144   &144   &144   &148   &148   &148   &148   \\ \cline{2-18}
				~                    &385   &149   &149   &148   &149   &149   &153   &154   &154   &154   &154   &157   &157   &157   &159   &159   &159   \\ \cline{2-18}
				~                    &401   &161   &159   &161   &165   &167   &166   &167   &167   &167   &167   &167   &167   &170   &170   &170   &172   \\ \cline{2-18}
				~                    &417   &172   &172   &173   &176   &176   &179   &179   &179   &179   &179   &180   &184   &184   &185   &185   &185   \\ \cline{2-18}
				~                    &433   &187   &188   &189   &189   &189   &189   &191   &191   &192   &196   &196   &196   &196   &197   &197   &197   \\ \cline{2-18}
				~                    &449   &201   &201   &204   &205   &204   &205   &205   &205   &205   &208   &210   &210   &210   &211   &211   &211   \\ \cline{2-18}
				~                    &465   &213   &215   &216   &216   &217   &217   &217   &218   &219   &219   &220   &222   &224   &225   &225   &229   \\ \cline{2-18}
				~                    &481   &227   &228   &230   &230   &230   &231   &233   &234   &235   &235   &236   &237   &239   &238   &242   &243   \\ \cline{2-18}
				~                    &497   &243   &243   &245   &247   &245   &248   &249   &248   &252   &253   &252   &253   &253   &254   &255   &256   \\ \cline{2-18}

				\hline
				
			\end{tabular}
		}
		\caption{Minus array for $L = 32, n =512$ and $1\le k\le n$. The value in the table is $\minus_{32}(n,k+\Delta)$.}
		\label{table:minus_n512}
	\end{table}
	
	\begin{table}[H]
		\centering
		\resizebox{\columnwidth}{!}{
			\begin{tabular}{c|c|cccccccccccccccc}
				\hline
				$n$ &\diagbox{$k$}{$\Delta$} &0 &1 &2 &3 &4 &5 &6 &7 &8 &9 &10 &11 &12 &13 &14 &15\\
				\hline
				\multirow{64}*{1024} &1     &0     &0     &0     &0     &0     &0     &0     &1     &0     &0     &1     &1     &3     &1     &1     &2     \\ \cline{2-18}
				~                    &17    &1     &1     &1     &1     &3     &1     &4     &5     &6     &5     &3     &1     &2     &3     &4     &2     \\ \cline{2-18}
				~                    &33    &1     &1     &1     &1     &1     &1     &1     &2     &2     &1     &1     &2     &3     &2     &1     &3     \\ \cline{2-18}
				~                    &49    &1     &2     &3     &3     &5     &3     &2     &3     &4     &5     &4     &3     &4     &3     &3     &4     \\ \cline{2-18}
				~                    &65    &5     &4     &5     &4     &5     &5     &5     &5     &4     &5     &3     &4     &5     &5     &6     &5     \\ \cline{2-18}
				~                    &81    &4     &4     &5     &6     &5     &6     &7     &5     &6     &7     &5     &5     &7     &5     &6     &7     \\ \cline{2-18}
				~                    &97    &5     &6     &6     &6     &6     &7     &7     &7     &8     &6     &7     &8     &9     &10    &10    &10    \\ \cline{2-18}
				~                    &113   &10    &10    &8     &9     &10    &10    &10    &10    &10    &10    &10    &10    &11    &11    &11    &10    \\ \cline{2-18}
				~                    &129   &11    &11    &11    &10    &10    &11    &10    &11    &12    &12    &12    &12    &12    &16    &16    &16    \\ \cline{2-18}
				~                    &145   &14    &12    &12    &16    &16    &16    &16    &16    &16    &16    &16    &16    &16    &16    &17    &17    \\ \cline{2-18}
				~                    &161   &18    &16    &17    &17    &17    &18    &17    &17    &16    &18    &19    &18    &18    &17    &18    &18    \\ \cline{2-18}
				~                    &177   &18    &18    &18    &17    &18    &19    &19    &19    &19    &20    &19    &20    &19    &20    &20    &21    \\ \cline{2-18}
				~                    &193   &20    &20    &21    &21    &20    &21    &20    &21    &20    &21    &22    &20    &21    &22    &22    &22    \\ \cline{2-18}
				~                    &209   &22    &21    &22    &22    &21    &22    &26    &26    &27    &26    &27    &28    &27    &29    &28    &29    \\ \cline{2-18}
				~                    &225   &28    &29    &27    &28    &29    &30    &31    &31    &31    &31    &32    &31    &34    &32    &31    &29    \\ \cline{2-18}
				~                    &241   &32    &31    &34    &33    &34    &32    &33    &34    &33    &34    &35    &36    &37    &38    &38    &38    \\ \cline{2-18}
				~                    &257   &38    &42    &42    &42    &42    &42    &42    &42    &42    &42    &42    &42    &40    &42    &42    &43    \\ \cline{2-18}
				~                    &273   &42    &42    &40    &42    &42    &43    &43    &45    &48    &48    &48    &46    &48    &48    &48    &49    \\ \cline{2-18}
				~                    &289   &48    &48    &48    &49    &51    &51    &49    &49    &49    &48    &49    &53    &51    &52    &53    &54    \\ \cline{2-18}
				~                    &305   &54    &53    &51    &53    &51    &54    &55    &57    &55    &54    &57    &57    &58    &59    &59    &60    \\ \cline{2-18}
				~                    &321   &58    &60    &64    &65    &65    &65    &63    &64    &65    &66    &64    &63    &65    &65    &64    &66    \\ \cline{2-18}
				~                    &337   &65    &66    &65    &65    &66    &66    &66    &64    &67    &66    &67    &66    &66    &64    &66    &67    \\ \cline{2-18}
				~                    &353   &69    &72    &72    &70    &72    &72    &72    &72    &72    &72    &76    &76    &75    &77    &79    &78    \\ \cline{2-18}
				~                    &369   &80    &80    &81    &80    &81    &80    &80    &81    &82    &83    &81    &80    &78    &81    &80    &79    \\ \cline{2-18}
				~                    &385   &79    &82    &81    &82    &80    &83    &82    &82    &86    &87    &87    &86    &84    &85    &89    &89    \\ \cline{2-18}
				~                    &401   &89    &88    &86    &87    &88    &89    &89    &89    &87    &86    &86    &87    &90    &91    &91    &91    \\ \cline{2-18}
				~                    &417   &93    &94    &95    &96    &96    &97    &97    &98    &100   &100   &102   &103   &102   &102   &100   &100   \\ \cline{2-18}
				~                    &433   &100   &100   &98    &96    &100   &98    &102   &100   &102   &102   &103   &103   &102   &103   &106   &107   \\ \cline{2-18}
				~                    &449   &108   &111   &111   &112   &112   &111   &111   &115   &116   &117   &116   &120   &120   &120   &122   &123   \\ \cline{2-18}
				~                    &465   &122   &121   &119   &120   &120   &122   &121   &122   &120   &121   &122   &123   &123   &125   &125   &123   \\ \cline{2-18}
				~                    &481   &124   &128   &128   &130   &128   &127   &128   &128   &130   &130   &130   &132   &135   &134   &136   &137   \\ \cline{2-18}
				~                    &497   &137   &137   &140   &140   &142   &143   &142   &141   &139   &140   &141   &142   &143   &143   &143   &143   \\ \cline{2-18}
				~                    &513   &146   &146   &147   &149   &147   &151   &152   &152   &152   &150   &152   &152   &150   &151   &151   &152   \\ \cline{2-18}
				~                    &529   &154   &155   &155   &154   &152   &152   &154   &154   &155   &155   &156   &159   &159   &159   &161   &161   \\ \cline{2-18}
				~                    &545   &160   &161   &162   &164   &164   &165   &166   &167   &168   &169   &169   &169   &167   &166   &164   &162   \\ \cline{2-18}
				~                    &561   &163   &164   &165   &166   &167   &167   &169   &169   &172   &172   &170   &168   &168   &169   &171   &172   \\ \cline{2-18}
				~                    &577   &172   &173   &175   &173   &175   &177   &176   &179   &178   &181   &182   &182   &182   &183   &183   &185   \\ \cline{2-18}
				~                    &593   &186   &184   &182   &183   &183   &182   &185   &186   &185   &189   &190   &188   &192   &190   &194   &194   \\ \cline{2-18}
				~                    &609   &196   &194   &195   &196   &197   &198   &201   &200   &201   &201   &205   &205   &205   &205   &203   &201   \\ \cline{2-18}
				~                    &625   &201   &203   &204   &205   &205   &205   &203   &204   &205   &205   &205   &207   &207   &208   &209   &207   \\ \cline{2-18}
				~                    &641   &207   &209   &212   &213   &212   &213   &213   &213   &213   &215   &215   &218   &220   &220   &220   &220   \\ \cline{2-18}
				~                    &657   &222   &223   &223   &225   &225   &223   &221   &222   &223   &224   &225   &225   &226   &227   &229   &230   \\ \cline{2-18}
				~                    &673   &230   &231   &231   &234   &235   &235   &237   &237   &238   &239   &241   &241   &242   &244   &244   &244   \\ \cline{2-18}
				~                    &689   &244   &244   &242   &241   &244   &245   &244   &245   &246   &246   &248   &249   &250   &251   &252   &253   \\ \cline{2-18}
				~                    &705   &253   &255   &255   &255   &255   &259   &259   &259   &262   &262   &262   &262   &265   &266   &266   &266   \\ \cline{2-18}
				~                    &721   &267   &267   &265   &265   &266   &267   &267   &268   &270   &268   &272   &273   &273   &274   &273   &276   \\ \cline{2-18}
				~                    &737   &276   &276   &278   &281   &281   &281   &284   &284   &284   &284   &285   &283   &283   &284   &285   &285   \\ \cline{2-18}
				~                    &753   &286   &285   &289   &289   &291   &291   &291   &291   &291   &292   &296   &296   &296   &296   &297   &296   \\ \cline{2-18}
				~                    &769   &296   &297   &298   &296   &296   &297   &297   &296   &297   &298   &301   &303   &304   &305   &305   &305   \\ \cline{2-18}
				~                    &785   &305   &308   &308   &308   &308   &310   &311   &311   &315   &315   &315   &316   &317   &320   &319   &320   \\ \cline{2-18}
				~                    &801   &321   &321   &321   &324   &327   &326   &327   &327   &327   &327   &327   &329   &329   &327   &329   &329   \\ \cline{2-18}
				~                    &817   &329   &329   &332   &332   &333   &335   &335   &335   &336   &338   &338   &342   &343   &342   &343   &343   \\ \cline{2-18}
				~                    &833   &343   &346   &347   &347   &347   &347   &347   &347   &351   &352   &352   &356   &356   &355   &356   &356   \\ \cline{2-18}
				~                    &849   &356   &358   &356   &356   &356   &356   &360   &364   &364   &362   &362   &364   &364   &364   &364   &368   \\ \cline{2-18}
				~                    &865   &369   &371   &371   &372   &374   &374   &374   &374   &374   &378   &379   &379   &380   &380   &380   &380   \\ \cline{2-18}
				~                    &881   &383   &384   &384   &384   &384   &387   &387   &388   &389   &389   &391   &391   &394   &394   &394   &394   \\ \cline{2-18}
				~                    &897   &393   &397   &399   &399   &402   &402   &402   &402   &402   &402   &406   &410   &409   &410   &409   &410   \\ \cline{2-18}
				~                    &913   &412   &412   &412   &412   &415   &415   &415   &418   &417   &418   &418   &417   &421   &423   &424   &426   \\ \cline{2-18}
				~                    &929   &426   &426   &427   &425   &426   &430   &432   &432   &430   &431   &432   &432   &434   &435   &436   &437   \\ \cline{2-18}
				~                    &945   &438   &438   &438   &438   &438   &440   &444   &445   &444   &446   &446   &445   &447   &448   &451   &453   \\ \cline{2-18}
				~                    &961   &453   &453   &456   &456   &457   &456   &457   &459   &459   &459   &461   &461   &463   &463   &466   &466   \\ \cline{2-18}
				~                    &977   &466   &468   &469   &468   &472   &471   &475   &473   &474   &475   &478   &476   &478   &478   &481   &480   \\ \cline{2-18}
				~                    &993   &482   &482   &486   &485   &486   &487   &489   &488   &489   &491   &491   &492   &493   &494   &495   &496   \\ \cline{2-18}
				~                    &1009  &497   &498   &499   &500   &501   &502   &503   &504   &505   &506   &507   &508   &509   &510   &511   &512   \\ \cline{2-18}

				\hline
				
			\end{tabular}
		}
		\caption{Minus array for $L = 32, n =1024$ and $1\le k\le n$. The value in the table is $\minus_{32}(n,k+\Delta)$.}
		\label{table:minus_n1024}
	\end{table}

	\bibliographystyle{IEEEtran}
	\bibliography{DPPolar}
	
\end{document}